# cyber justice

## LA TRANSFORMATION NUMÉRIQUE DU DROIT ET DE LA JUSTICE

# La transformation numérique de la justice

## Ambitions, réalités et perspectives

Étude réalisée sous la direction de Yannick Meneceur, maître de conférences associé à l'Université de Strasbourg et expert associé à l'Institut Robert Badinter

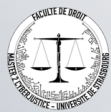
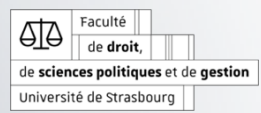
Faculté de droit, de sciences politiques et de gestion
Université de Strasbourg

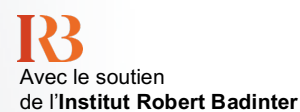
Avec le soutien de l'Institut Robert Badinter

# La transformation numérique de la justice
## Ambitions, réalités et perspectives

Étude réalisée sous la direction de Yannick Meneceur, magistrat[1], maître de conférence associé à l'Université de Strasbourg et expert associé à l'Institut Robert Badinter.

Ces travaux s'inscrivent dans le cadre des enseignements du Master Cyberjustice de la Faculté de droit, de sciences politiques et de gestion de l'Université de Strasbourg, avec le concours des étudiants de M2[2] (promotions 2022-2023, 2023-2024, 2024-2025 et 2025-2026).

Le directeur scientifique et les étudiants tiennent à remercier l'ensemble des contributeurs, ainsi que les professeurs Catherine Ledig, Bénédicte Girard et Emmanuel Netter, pour leur soutien dans la réalisation de cette étude.

---

[1] Les analyses et opinions exprimées n'engagent pas le ministère de la justice, mais seul le directeur scientifique en sa qualité de chercheur associé.
[2] Les noms des étudiants contributeurs à cette étude sont en annexe 2.

# Présentation de l'étude

En France, l'emploi de l'informatique comme levier d'amélioration de l'efficacité de la justice[3] s'inscrit dans une stratégie de modernisation datant des années 1960 avec l'emploi des premiers calculateurs.

Après avoir centralisé et automatisé la mémoire de la justice pénale dès le début des années 1980 avec le Casier judiciaire national, industrialisé progressivement la production de documents dans les tribunaux et remplacé les registres papiers par des bases de données pour le suivi électronique des affaires, de nouvelles technologies recomposent maintenant en profondeur l'offre de justice en rendant possible une dématérialisation totale et sécurisée des processus (blockchains), la résolution de litiges en ligne, des analyses avancées de la jurisprudence ou encore la synthèse des pièces d'une affaire.

Le secteur privé, avec de jeunes pousses innovantes (les legaltechs), se présente comme pionnier de l'innovation, en s'appropriant avec pragmatisme et enthousiasme les outils de la justice numérique du XXIe siècle. En contraste, les magistrats et greffiers dans les juridictions dénoncent régulièrement de leur côté l'obsolescence de leurs propres systèmes d'informations et les difficultés à capitaliser les bonnes pratiques. Entre initiatives locales et emploi discret d'offres du commerce (« shadow IT » et « shadow AI »), les acteurs de terrain demeurent toutefois prompts à se saisir d'outils leur permettant de répondre à une charge de travail, toujours plus croissante.

La présente étude[4], réalisée sur un cycle de quatre années universitaires avec le concours des étudiants du M2 du Master Cyberjustice de la Faculté de droit, de sciences politiques et de gestion de l'Université de Strasbourg, vise à objectiver les discours et les représentations de la transformation numérique de la justice, notamment au travers de la capitalisation de témoignages de professionnels du domaine[5] et d'une exploitation de la littérature disponible.

---

[3] Le terme de « justice » sera à entendre dans le cadre de cette étude de manière restreinte, pour ne traiter que de la transformation numérique des tribunaux et des professions du droit concourant à l'activité judiciaire. Il ne sera pas traité de l'informatisation des services déconcentrés du ministère de la justice, c'est-à-dire de l'administration pénitentiaire et de la protection judiciaire de la jeunesse.

[4] Cette étude s'est appuyée pour son cadrage sur une précédente publication du directeur scientifique : Y. Meneceur, « La transformation numérique de la justice – Ambitions, réalités et perspectives », *Les Cahiers Français*, n°416, juillet-août 2020

[5] Les noms des professionnels entendus dans le cadre de ce cycle de travaux sont en annexe 1. Les interventions des professionnels cités ont été réalisées en leur capacité personnelle, au regard de leur expérience passée ou présente d'informatisation et de numérisation des services judiciaires. Les analyses et opinions exprimées dans l'étude ne les engage ni eux, ni leurs institutions d'appartenance.

# Sommaire

**Introduction**

1. **La transformation numérique de la justice, « cœur du réacteur » de la justice du XXIe siècle** – En 2018, la transformation numérique de la justice était désignée par Nicole Belloubet, ministre de la justice, comme le « cœur du réacteur » de la justice du XXIe siècle car déterminante pour la suite de l'ensemble des réformes de la justice[6]. La ministre présentait alors les ambitions gouvernementales traduites dans la loi du 23 mars 2019 de programmation 2018-2022 et de réforme pour la justice (LPJ). De manière inédite, 530 millions d'euros et 260 emplois supplémentaires étaient alloués à l'informatique judiciaire en vue, notamment, de dématérialiser les procédures et de créer un véritable service public numérique de la justice[7].

Malgré un engagement actif de l'administration centrale de la justice dans de nombreux chantiers de transformations, comme la procédure pénale numérique (PPN), la Cour des comptes dressait en janvier 2022 un sévère bilan critique de ce plan de transformation numérique (PTN) à l'occasion d'un point d'étape réalisé à la demande de la Commission des finances du Sénat[8]. Pour les magistrats financiers, ce plan semblerait tout d'abord plutôt qualifiable de plan de rattrapage, au vu des importants retards structurels constatés. L'insuffisance du renforcement de la fonction informatique du ministère, des choix présentés comme contestables dans les priorités des projets et un manque de suivi budgétaire sont également soulignés dans ce rapport. En 2023, une mission interministérielle confiée à l'inspection générale de la justice, l'inspection générale des finances et le Conseil général de l'économie, de l'industrie, de l'énergie et des technologies a également mené un travail pour évaluer, de manière transversale, la capacité du ministère de la justice à piloter une stratégie de transformation numérique globale[9].

C'est dans ce contexte que le ministre de la justice Éric Dupond-Moretti annonçait un nouveau PTN le 14 février 2023, en s'appuyant notamment sur les conclusions « Le numérique pour la justice » des États généraux de la justice[10] et en désignant trois nouvelles priorités : un objectif « zéro papier » d'ici 2027, l'arrivée de « vrais informaticiens » en juridiction et le lancement d'une application, portail avec les justiciables[11].

Sans remettre en cause cette trajectoire, Gérald Darmanin déclarait à l'occasion de la remise d'un rapport sur « l'IA » et la justice[12] en 2025 que « les opportunités créées par l'IA pour améliorer l'efficacité et la qualité de la Justice au bénéfice des magistrats,

---

[6] M. Babonneau, « Chantiers de la justice : la transformation numérique, 'cœur du réacteur' », *Dalloz Actualité*, 16 janvier 2018
[7] L'ensemble des actions en cours du ministère de la justice sur la transformation numérique sont accessibles en ligne : https://www.cours-appel.justice.fr/nancy/la-transformation-numerique-du-ministere-de-la-justice, consulté le 19 février 2023
[8] Améliorer le fonctionnement de la justice – Point d'étape du plan de transformation numérique du ministère de la justice, Communication à la Commission des finances du Sénat, janvier 2022
[9] Ce rapport n'est pas public et ses conclusions ne peuvent être présentées dans la présente étude.
[10] Annexe 25, Rapport sur le numérique du comité de pilotage des États généraux de la justice, 17 mars 2022
[11] A. Mestre, « Éric Dupond-Moretti présente son 'plan de transformation numérique' pour la justice », *Le Monde*, 14 février 2018
[12] Rapport L'IA au service de la justice : stratégie et solutions opérationnelles, mai 2025, accessible sur : https://www.justice.gouv.fr/sites/default/files/2025-08/rapport_ia_au_service_de_la_justice.pdf, consulté le 15 novembre 2025

agents et usagers sont immenses ». La constitution d'une direction de programme IA et d'un observatoire de haut niveau dédié était lancée dans la foulée de la remise du rapport.

La dernière décennie des politiques publiques en matière informatique dans les tribunaux judiciaires se caractérise donc par une succession d'annonces et d'orientations pour doter les services judiciaires d'outils informatiques au niveau des standards contemporains et des moyens inédits, mais encore trop modestes pour produire des résultats tangibles. Les très nombreux rapports de bilan produits durant cette période convergent sur ces points[13].

2. **L'innovation de la justice poussée par le secteur privé** – En total contraste avec une initiative publique peinant à produire ses pleins effets, c'est du côté du secteur privé que nombre de décideurs publics vont chercher des exemples en ce qui concerne l'innovation dans le domaine de la justice et du droit. En France, depuis 2016 et sur le modèle de leurs aînés de la Silicon Valley, les jeunes pousses spécialisées dans le domaine du droit (les *legaltechs*) ont su ménager leurs effets d'annonces et créer de grands espoirs.

L'ambition de ces entrepreneurs est venue dépasser, en effet, la simple offre habituelle de gestion ou de base de données : elle visait, de manière inédite, à apporter une valeur ajoutée sur l'interprétation du droit et sa mise en œuvre, en s'appuyant sur des technologies de rupture telles que l'intelligence artificielle (« IA[14] », notamment l'apprentissage automatique[15]). Ces entreprises ont semblé en capacité d'en tirer rapidement parti, et de manière plus agile que le secteur public, en les habillant d'une offre inédite de services et, surtout, d'une narration de modernité. Et même si cette offre, parfois aux fondements approximatifs, répondait en réalité aux besoins des professions juridiques du secteur privé comme les cabinets d'avocats, les directions juridiques ou les compagnies d'assurance, une forme de confusion a nourri des discours ambitionnant d'importer, au prétexte de cette même modernité, ces outils dans les tribunaux[16]. Cette précipitation a d'ailleurs déjà conduit à de grandes

---

[13] Il pourrait y être ajouté le rapport de l'Institut Montaigne, Justice : Faites entrer le numérique, novembre 2017, accessible sur : https://www.institutmontaigne.org/publications/justice-faites-entrer-le-numerique, consulté le 19 février 2023 ou le rapport des chantiers de la justice « Transformation numérique », janvier 2018, accessible sur : http://www.justice.gouv.fr/publication/chantiers_justice/Chantiers_justice_Livret_01.pdf, consulté le 19 février 2023
[14] L'acronyme d'intelligence artificielle sera présenté entre guillemets par commodité éditoriale. L'ensemble des technologies recouvertes par ce terme ne constituent naturellement pas une personnalité autonome et, afin de se garder de tout anthropomorphisme, il a été choisi de résumer les termes plus appropriés « d'outils d'intelligence artificielle » ou « d'applications d'intelligence artificielle » par le seul terme « d'IA » entre guillemets.
[15] Pour une revue des domaines d'application de l'IA par les legaltechs, V. Par exemple R. Slama, A. Louis, V. Callipel (sup.), « Étude comparative d'outils d'intelligence artificielle offerts par les legaltechs aux professionnels du droit », *Laboratoire de Cyberjustice, projet ACT*, mars 2023, accessible sur : https://cyberjustice.openum.ca/files/sites/102/Projet_Legal-Startups-FINAL.pdf, consulté le 4 mars 2023
[16] B. Louvel, allocution lors du colloque « La justice prédictive », 12 février 2018, accessible sur : https://www.courdecassation.fr/toutes-les-actualites/2018/02/12/bertrand-louvel-la-justice-predictive, consulté le 29 novembre 2025

déceptions[17], comme en témoigne l'engouement, aujourd'hui apaisé, autour de l'emploi de jurimétrie[18].

Malgré un ralentissement en 2022 dû à la concentration des acteurs[19], une recomposition en profondeur de l'offre de justice et des métiers y concourant est bien toujours à l'œuvre, avec le développement notamment d'une dématérialisation totale des processus (avec l'appui de blockchains), la résolution de litiges en ligne, des analyses statistiques avancées de la jurisprudence et l'emploi de modèles génératifs pour créer et analyser des documents. Ces outils viennent opportunément au soutien de politiques publiques européennes favorisant la résolution de contentieux hors des tribunaux par des professionnels spécialisés (médiateurs, conciliateurs et arbitres), pour désengorger des juridictions saisies d'un nombre toujours plus croissant d'affaires[20]. La justice est ainsi devenue nouvel objet de marché, avec une offre privée, dont la symbolique et les codes empruntent parfois au secteur public[21].

3. **Aller au-delà des représentations de professionnels du droit et de la justice rétifs au changement** – Ce très rapide état des lieux, nécessairement incomplet et ne restituant pas toutes les nuances et la complexité de la situation étudiée, permet toutefois d'illustrer la ténacité de certaines représentations comme celle d'une prétendue résistance chronique au changement des professionnels de la justice, contrastant avec l'agilité de jeunes « startupers » à qui il conviendrait de donner tous les moyens d'agir[22].

Pourtant, l'informatique publique judiciaire a été florissante à la fin des années 1980 : quelques agents des greffes et magistrats, passionnés de ces nouveaux objets techniques, ont rapidement vu l'opportunité d'industrialiser des tâches, comme la mise en forme des jugements. Le passage de la machine à écrire électrique aux ordinateurs personnels a été rapidement perçu comme une évolution tout à fait naturelle de l'outillage et l'informatique d'initiative locale a permis aux tribunaux d'être parmi les administrations françaises les mieux informatisées au début des années 1990.

4. **D'un patrimoine à une dette au sein des tribunaux** – Mais ce qui était un patrimoine semble aujourd'hui s'être transformé en dette, la France se situant en 2022-2023 dans le dernier tiers des évaluations des systèmes judiciaires européens[23]. Nombre de logiciels développés dans les années 1990 et au début des années 2000 sont encore en fonctionnement pour des processus critiques, avec nombre d'aléas de maintenance

---

[17] L'utilisation de l'outil Predictice déçoit la cour d'appel de Rennes, Dalloz Actualité, 16 octobre 2017, accessible sur : https://www.dalloz-actualite.fr/interview/l-utilisation-de-l-outil-predictice-decoit-cour-d-appel-de-rennes, consulté le 29 novembre 2025

[18] Sur l'objectivation de la jurimétrie à contre-courant des discours de l'époque, V. Y.Meneceur, « avenir pour la justice prédictive ? – Enjeux et limites des algorithmes d'anticipation des décisions de justice », La Semaine Juridique Edition Générale n°7, 12 Février 2018

[19] V. le baromètre des legaltechs, édition 2022, réalisé par Lamy, Maddyness et la Banque des territoires, accessible sur : https://www.maddyness.com/app/uploads/2023/01/BAROMETRE_LEGALTECH_2023.pdf, consulté le 19 février 2023

[20] Voir par exemple les séries statistiques publiées sur le site de la CEPEJ, accessibles sur : https://public.tableau.com/app/profile/cepej/viz/OverviewFR/Overview, consulté le 20 février 2023

[21] Ainsi l'actuel Centre d'arbitrage des affaires familiales avait été lancé sous le nom de tribunal arbitral des affaires familiales, avec un visuel présentant un palais de justice.

[22] V. par exemple A. van den Branden, *Les robots à l'assaut de la justice*, Larcier, 2019

[23] Cf. infra et v. de manière générale les indices de la CEPEJ sur l'emploi des technologies de l'information, accessible sur : https://public.tableau.com/app/profile/cepej/viz/ICTFR/Development, consulté le 20 février 2023

et de sécurité. Les tentatives d'innovation du terrain sont aujourd'hui portées avec des modalités comme les EIG (entrepreneurs d'intérêt général) et le soutien d'un incubateur au ministère de la justice[24]. Mais, durant la première partie des années 2020, les résultats tangibles à grande échelle peinent à émerger et ne permettant pas d'inverser la narration négative sur l'informatique judiciaire, émanant d'un grand nombre de magistrats et fonctionnaires : parmi les griefs communément soulevés par les utilisateurs des juridictions, il peut être cité le manque de performance des réseaux informatiques, la multiplicité d'applications (parfois obsolètes) ne communiquant pas entre elles, ou encore une hiérarchisation contestée des besoins.

Les raisons de cette situation n'ont pas manqué d'être analysées et des critiques sévères régulièrement relayées dans la presse[25], sur fond d'un manque récurrent de moyens dénoncé par les acteurs de terrain. Aux risques classiques de mécanisation de la justice s'ajoutent également de la part de responsables syndicaux des critiques de fond sur les dérives des politiques de nouvelle gestion publique (*new public management*), dont l'informatique serait l'outil opérationnel.

Le développement même de nouvelles solutions se heurte, de plus, à un environnement réglementaire mouvant : outre les dispositions relatives à la protection des données, le règlement sur l'intelligence artificielle adopté par l'Union européenne en 2024 place dans la catégorie de systèmes à haut risque certaines des applications relatives à l'administration de la justice. Des exigences de stricte mise en conformité vont donc progressivement s'imposer tant au secteur public qu'au secteur privé dans les années à venir, avec une articulation parfois complexe à trouver avec les cadres juridiques existants, comme ceux relatifs à la protection des données[26].

L'objectivation des réussites et des échecs de la transformation numérique de la justice est donc ardue : alors même que l'effort et l'engagement des services d'administration centrale de la justice au niveau opérationnel se sont incontestablement accrus, ces mêmes services semblent freinés par de multiples causes, aggravées par la forte inertie propre à toutes les grandes macrostructures administratives. Le « paradoxe de Solow », constatant à la fin des années 1980 l'absence de gains de productivité dans les organisations malgré une généralisation de l'informatisation, semblerait encore d'une particulière pertinence pour caractériser la situation[27]. Il n'y a eu toutefois que

---

[24] Le ministère de la justice a aussi son incubateur !, Village de la justice, janvier 2023, accessible sur : https://www.village-justice.com/articles/incubateur-ministere-justice,44735.html, consulté le 19 février 2023

[25] V. notamment P. Gonzalès, « Le grand bazar de l'informatique judiciaire », *Le Figaro*, 4 juin 2020, ou J.-B. Jacquin, « La crise sanitaire met en lumière la faiblesse de l'institution judiciaire », *Le Monde*, 7 juin 2020 ou A. Vidalie, « Serveurs inaccessibles, logiciels archaïques… Les bugs de la machine judiciaire », *L'Express*, 8 juin 2020 ou, enfin, G.Thierry, « Le confinement, crash test de la transformation numérique de la justice », *Dalloz Actualité*, 10 juin 2020

[26] Dans le temps de finalisation de cette étude, la publication par la Commission européenne d'un projet de réforme de la législation européenne numérique, portée dans un « omnibus » numérique, laisse supposer un report de certaines dispositions du règlement sur l'intelligence artificielle et des modifications de certaines dispositions du règlement général sur la protection des données.

[27] Pour profiter pleinement des technologies de l'information, les entreprises devraient entreprendre une refonte systématique de leurs méthodes de travail. Pour les arguments en ce sens, V. C. Chamaret, « En finir avec le paradoxe de Solow », *Sociétal*, n°68, 2010

très peu de recherches approfondies pour explorer ces hypothèses et mesurer précisément les effets de l'informatisation sur l'efficacité et la qualité de la justice[28].

5. **Plan** – En préalable à de telles démarches, cette étude visera principalement à documenter, au moins partiellement, l'historique de la transformation numérique de la justice, ses ambitions et ses principales réalisations (I), à mesurer la réelle portée de cette transformation (II) et à dresser les perspectives d'évolutions déjà perceptibles au mitan des années 2020 (III).

---

[28] V. par exemple Conseil de l'Europe, Systèmes judiciaires européens, efficacité et qualité de la justice – Rapport thématique : l'utilisation des technologies de l'information dans les tribunaux en Europe, Études de la CEPEJ n°24, décembre 2016

I. **De l'informatique judiciaire à la transformation numérique de la justice : les étapes d'une ambition de rationalisation et d'efficacité**

I.1. **Avant 1980 : l'informatique s'implante dans la justice avec des ordinateurs centraux (mainframes)**

I.1.1. *Contexte de l'introduction de l'informatique au sein du ministère de la justice*

6. **Les prémices de l'informatisation de la justice** – L'administration de la justice en France a investi très tôt l'emploi de l'informatique. Ainsi, le Centre de recherche de l'éducation surveillée à Vaucresson a été certainement le premier établissement de cette administration à se doter d'un calculateur en 1966[29].

Dans l'élan du « plan Calcul »[30], le ministère de la justice a été également l'une des premières administrations à se doter, par un arrêté du 13 février 1967, d'une « Commission de l'informatique[31] » en recevant pour mission « d'étudier les moyens scientifiques, administratifs et financiers qui permettraient d'appliquer les méthodes de traitement de l'information aux problèmes de la documentation juridique, de la police judiciaire, des statistiques, et d'une manière générale aux problèmes de l'administration de la Justice ».

Mais déjà, les professionnels de terrain, notamment des magistrats[32], exprimaient de la méfiance vis-à-vis de ces nouveaux outils, avançant notamment des risques théoriques de dénaturation de l'acte de juger et de la justice.

I.1.2. *L'exemple pionnier du Casier judiciaire national*

7. **Contexte de l'informatisation du Casier judiciaire national** – L'emploi d'ordinateurs centraux pour l'administration de la justice ne sera toutefois pas compromis par les réticences des professions. Le lancement d'études d'informatisation du Casier judiciaire national commencera dès 1973 et le système informatique sera concrétisé en 1980 avec la loi du 4 janvier relative à l'automatisation du casier judiciaire, suivie d'un décret du 6 novembre 1981. Ces dispositions ont alors autorisé, avec l'aval de la CNIL, la tenue d'un fichier sur le passé pénal des individus par un service national placé sous l'autorité du ministre de la justice[33].

---

[29] J. Verin, « La commission de l'informatique du ministère de la justice », *Revue internationale de droit comparé*, Vol. 20 n°4, octobre – décembre 1968, pp. 673-676

[30] Ce plan de 1967 visait alors à assurer l'indépendance technologique de la France en matière informatique afin de ne pas dépendre des constructeurs américains.

[31] Intervention de Jean-Pierre Poussin le 23 septembre 2022. La « Commission de l'informatique, des réseaux et de la communication électronique » (COMIRCE) lui a succédé en 1996 : Arrêté du 5 juin 1996 relatif relatif à la commission de l'informatique, des réseaux et de la communication électronique du ministère de la justice, NOR : JUSA9600173A, JORF n°135 du 12 juin 1996, accessible sur : https://www.legifrance.gouv.fr/jorf/id/JORFTEXT000000560330, consulté le 18 février 2023. Cette Commission a été absorbée par la création du secrétariat général du ministère de la justice en 2008.

[32] Voir par exemple M. Ancel, « Les problèmes posés par l'application des techniques nouvelles au droit pénal et à la procédure pénale », *Rassegna di profilassi criminale ed psichiatrica*, 1968, pp. 3-22, cité par J. Verin, *op. cit.* ou J.-P. Buffelan, « L'informatique judiciaire au ministère français de la justice », *Informat. e diritto*, 1977, pp. 63 – 91.

[33] Ch. Elek, *Le Casier judiciaire*, Que sais-je ?, PUF, 1988 et Histoire du casier judiciaire national, site internet du ministère de la justice, 2009 : http://www.archives-judiciaires.justice.gouv.fr/index.php?article=15012&rubrique=10774&ssrubrique=10828

8. **Histoire de la création d'un Casier judiciaire national** – Spécifiquement, l'histoire du Casier judiciaire national montre une construction progressive d'un outil pensé à la fois pour la gestion pénale et pour l'organisation de l'État. Institué au XIXᵉ siècle afin de permettre un suivi centralisé des condamnations, il répondait initialement à des enjeux de contrôle électoral et de lutte contre la récidive. Son fonctionnement reposait d'abord sur des fichiers papier répartis entre les tribunaux, puis sur un casier central chargé de regrouper les informations concernant les personnes nées à l'étranger. L'augmentation du volume des données, notamment après les indépendances, a rendu nécessaire une modernisation profonde et a conduit, dans les années 1970, à la mise en place des premières solutions informatiques pour l'identification et la gestion des fiches personnelles.

9. **Les étapes de l'informatisation du Casier judiciaire national** – L'informatisation a d'abord porté sur la recherche d'identité, en raison de la complexité des données recueillies pour les personnes nées à l'étranger, puis sur la gestion des condamnations elles-mêmes. La transition vers un casier centralisé a entraîné la nécessité de saisir et de contrôler manuellement un grand nombre d'informations, ce qui a conduit au développement de procédures de double saisie, de contrôles de cohérence et d'outils juridiques automatisés afin de prévenir les erreurs. L'intégration des données d'état civil provenant du répertoire national d'identification des personnes physiques a constitué une autre étape structurante, permettant une harmonisation accrue des identités enregistrées, malgré les risques persistants d'erreur liés aux multiples étapes de copie et de transcription.

Avec l'essor des échanges électroniques, le Casier judiciaire national est devenu un système centralisé capable de communiquer à distance avec les juridictions et les administrations, tout en maintenant un haut niveau de contrôle interne et de traçabilité. L'automatisation a également transformé les usages : les demandes de bulletins pour les particuliers se sont dématérialisées, les interconnexions européennes se sont développées, et le fichier s'est ouvert à de nouveaux registres spécialisés, notamment pour les infractions sexuelles ou le terrorisme. Cette évolution s'inscrit dans une logique d'efficacité administrative, mais aussi de rigueur juridique, car le Casier judiciaire demeure un outil où la précision des données conditionne directement les droits et libertés des personnes concernées, ce qui explique l'attention constante portée à la qualité des informations et à la sécurité des systèmes qui les hébergent.

Le Casier judiciaire national poursuit aujourd'hui un projet de dématérialisation totale pour la délivrance de l'ensemble des bulletins (projet ASTREA – Application de Stockage, de Traitement et de Restitution des Antécédents judiciaires). Ce projet porte sur la refonte du système d'information du Casier judiciaire national et vise notamment à moderniser les applications obsolètes de gestion des condamnations (pour les personnes physiques et morales), à ouvrir l'accès à des extraits de casier judiciaire en continu (24h/24–7j/7) et à interconnecter ce système avec les casiers judiciaires européens. Le palier 2 du projet (consacré au casier judiciaire des personnes morales) a été mis en service en janvier 2022, tandis que le palier 3, relatif à l'enregistrement, la gestion et la restitution des décisions à l'encontre des personnes physiques, est toujours en cours de réalisation. La loi d'orientation et de programmation de la justice

de 2023 rappelle que l'achèvement de la modernisation du casier judiciaire national sera assuré par la finalisation d'ASTREA et du volet européen ECRIS-TCN[34].

**I.2. Entre 1980 et 2000 : développement de la micro-informatique, de la bureautique et des systèmes de gestion d'affaires en s'appuyant sur l'initiative locale**

10. **Les opportunités offertes par le développement de la micro-informatique** – Sous l'impulsion de la miniaturisation des ordinateurs et de leur développement industriel (micro-informatique), la deuxième phase de la transformation numérique de justice, que l'on pourrait situer approximativement entre les années 1980 à 2000, s'est concentrée sur l'industrialisation de la production de documents et l'automatisation de la gestion des affaires, en y incluant la fiabilisation de la production statistique. La poursuite de cet élan se fondra progressivement dans les politiques interministérielles de coordination de l'e-administration, qui peuvent être datées du PAGSI (programme d'action gouvernemental pour la société de l'information) en 1998, puis du plan ADELE (ADministration ELEctronique) entre 2004 et 2007. L'agence pour le développement de l'administration électronique (ADAE), créée en 2003 et fondue depuis dans la DINUM (direction interministérielle du numérique), avait assurée la mise en œuvre de ce plan[35].

11. **Le rôle des fonctionnaires et des magistrats dans l'informatisation des juridictions** – Les fonctionnaires de greffe ont été les principaux acteurs opérationnels de cette phase et se sont appropriés dès les années 1980 la micro-informatique, dans un contexte où s'exprimaient déjà de forts ressentis de manques de moyens et de personnels. Quelques expérimentateurs, le plus souvent sur leur propre initiative et avec le soutien local de leur juridiction et de quelques magistrats, vont ainsi faire passer leurs tribunaux des machines à écrire électriques et des registres papiers vers un nouvel outil : les micro-ordinateurs. Cette informatique dite « d'initiative locale », qui a consisté à l'utilisation directe par les juridictions de leurs crédits déconcentrés pour procéder à l'informatisation de leurs services, a permis à dans les années 1990 d'accélérer significativement l'appropriation par les juridictions de ces technologies[36], avec un appui de l'administration centrale pour déployer localement des serveurs et des applications (micro et mini pénale en matière pénale, chaînes WinCi en matière civile), en plus d'offrir de nouveaux services (informatisation de la gestion du personnel des greffes et de la magistrature, intranet justice). Les ordinateurs centraux (*mainframes*) se déploieront plutôt en région parisienne en matière pénale (NCP – nouvelle chaîne pénale) ou pour des besoins centraux (gestion des ressources humaines).

Quelques magistrats ont également pu aussi investir ces politiques d'initiative locale. Le juge Victor-Amédée Dieuzaide, juge au tribunal d'instance de Marseille, a ainsi créé

---

au milieu des années 1980 ce qui devait être l'une des toutes premières chaînes informatiques locales dans ce tribunal. Alain Nuée, maintenant premier président honoraire, a expérimenté à Nancy une mise en état civile par minitel dans les années 1990. Thierry Ghera a mis en place une mise en état par voie électronique en 1998 au tribunal de grande instance d'Alès, appuyée par une convention étant signée avec le barreau local. Toujours pour la mise en état, le système « Melgreffe » a aussi été développé à Nancy et Nîmes.

## I.3.  À partir des années 2000 : la centralisation des principaux systèmes d'information justice

### I.3.1.  Contextualisation de la centralisation des systèmes d'information de la justice

12. **La vision du « tribunal du futur » au début des années 2000** – Un projet de « tribunal du futur » sera présenté à Caen le 1er décembre 2003 avec l'appui de la COMIRCE, en étudiant notamment les conséquences de l'emploi de la visioconférence sur la procédure pénale.

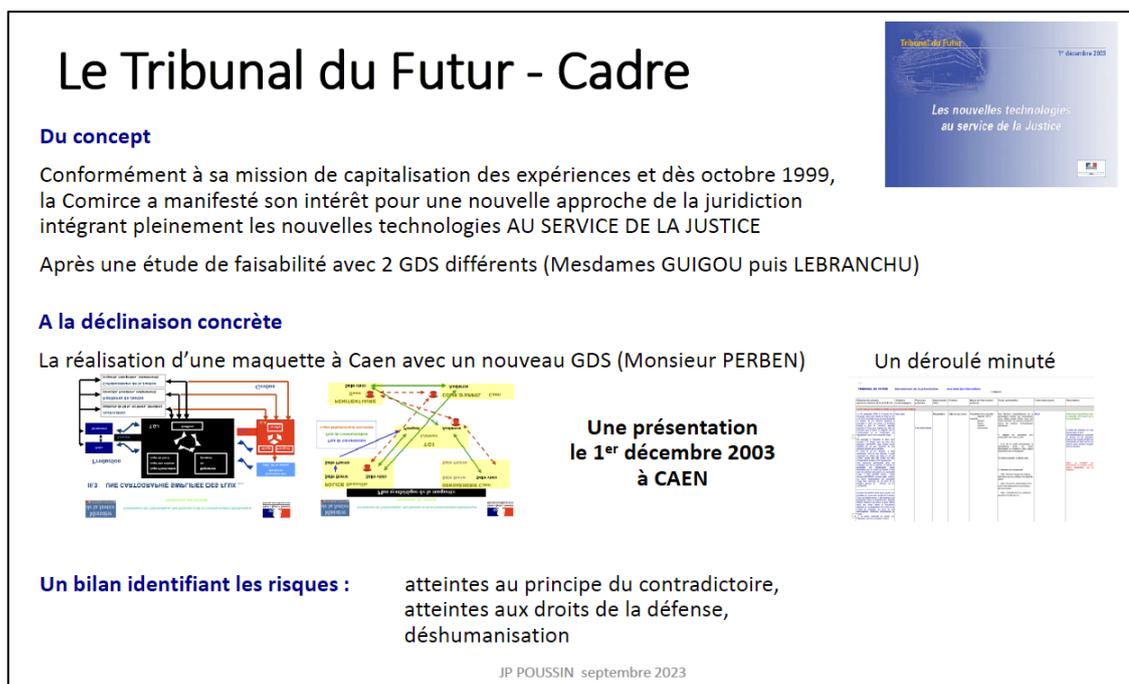

Source : Jean-Pierre Poussin, intervention du 23 septembre 2022 devant le M2 Cyberjustice

Mais l'essentiel des efforts de l'informatique judiciaire a plutôt porté au début des années 2000 sur une rationalisation et une centralisation de la conception des applications de gestion, notamment en matière pénale.

13. **Une centralisation motivée par des exigences de rationalisation** – Si les politiques de décentralisation des crédits informatiques dans les années 1980 ont produit un effet d'accélération, elles ont conduit dans le même temps à une très grande hétérogénéité des systèmes et ont complexifié leur gestion, chaque juridiction devant se doter de compétences spécialisées pour assurer le bon fonctionnement des serveurs et ordinateurs. D'autres initiatives, par ailleurs, n'ont tenu qu'à la personnalité de leur promoteur et n'étaient pas aisément maintenables ou industrialisables.

Afin de simplifier la maintenance et de rationaliser les coûts, le pilotage stratégique et la mise en œuvre opérationnelle des opérations d'informatisation ont donc été centralisés au ministère de la justice au début des années 2000, en plus de consolider la gestion de serveurs (comme ceux de courrier électronique).

14. **Une centralisation motivée en appui à des réformes** – Il doit aussi être rappelé que l'informatique judiciaire a servi de levier à de nombreuses réformes de fond. Ainsi, CASSIOPEE a été l'outil de la réforme de la carte judiciaire en matière pénale en permettant la fusion de base de données locales par un transfert vers une base nationale. C'est pourquoi son lancement a été réalisé de manière très volontariste en 2008, malgré une base technique qui était encore en cours de stabilisation. La communication électronique civile, étendue aux juridictions de seconde instance, a été aussi accompagné de la décision de supprimer la profession d'avoué près les cours d'appel[37].

15. **Des perspectives se heurtant à l'héritage de l'informatique d'initiative locale et à des budgets contraints** – La modernisation informatique de gestion, tant en matière civile que pénale, s'est réalisée à un rythme volontaire, mais s'est heurtée à son propre héritage, devenu une véritable dette. Le ministère de la justice a cherché durant des décennies à concilier les exigences du temps politique et le maintien en conditions opérationnelles de l'existant, au gré de budgets contraints où l'informatique a longtemps servi de variable d'ajustement.

### I.3.2. En matière pénale

16. **CASSIOPEE** – C'est dans ce contexte que sont lancés des projets d'applications informatiques nationales tels que la chaîne pénale « CASSIOPEE[38] » qui visait alors à remplacer la multiplicité d'applications sectorielles nécessaires pour traiter des affaires correctionnelles. Il pourrait aussi être cité APPI pour l'application des peines et, du côté de l'administration pénitentiaire, GENESIS.

CASSIOPEE connaîtra d'importantes difficultés de déploiement à la fin des années 2000[39] et parviendra à finaliser une première phase dans l'ensemble des tribunaux de grande instance (maintenant tribunaux judiciaires) en mai 2013.

L'application constitue le bureau d'ordre national automatisé des procédures judiciaires prévu par l'article 48-1 du code de procédure pénale et ne supporte pas la dématérialisation des procédures. CASSIOPEE a, en revanche, intégré un module

---

[37] V. le bilan dressé par le Sénat sur la suppression de la profession d'avoué : Rapport d'information sur la mise en œuvre de la loi réformant la procédure d'appel, n°580, 4 juin 2014 accessible sur : http://www.senat.fr/rap/r13-580/r13-5801.pdf, consulté le 4 mars 2023

[38] Chaine Applicative Supportant le Système d'Information Orienté Procédure pénale Et Enfants

[39] Parmi les articles de presse à la suite du rapport parlementaire conduit par Etienne Blanc, v. par exemple J.M. Leclerc, « Le grand bug informatique de Cassiopée freine la justice », *Le Figaro*, 1er mars 2011, accessible sur : https://www.lefigaro.fr/actualite-france/2011/02/28/01016-20110228ARTFIG00708-le-grand-bug-informatique-de-cassiopee-freine-la-justice.php, consulté le 27 novembre 2025

d'intégration et d'export de données de procédures. Ainsi les échanges inter-applicatifs (EIA) ont été opérationnalisés avec la gendarmerie nationale (entre LRPGN et CASSIOPEE), puis avec la police nationale (entre LRPPN et CASSIOPEE), CASSIOPEE devant alimenter théoriquement en sortie le successeur unique des fichiers d'antécédents STIC (système de traitement des infractions constatées, police) et JUDEX (système judiciaire de documentation et d'exploitation, gendarmerie), le TAJ (traitement des antécédents judiciaires).

CASSIOPEE poursuivra un difficile déploiement dans les juridictions d'appel après 2020[40]. L'application est toujours fonctionnelle en première instance en 2025 et se trouve à nouveau sous étude d'un cabinet de conseil pour sa « simplification »[41].

17. **La procédure pénale numérique (PPN)** – La procédure pénale numérique (PPN), quant à elle, est un programme interministériel lancé en 2018 entre le ministère de la justice et le ministère de l'intérieur, avec le soutien politique direct du président de la République, annoncé lors de l'audience solennelle de rentrée de la Cour de cassation en 2018. Cette impulsion présidentielle a permis de donner au projet un caractère prioritaire, assorti d'exigences fortes de résultats et d'un calendrier ambitieux : la mise en place d'une procédure pénale nativement numérique et probante à l'horizon 2022.

Conçue comme un programme complémentaire à CASSIOPEE, la PPN vise à permettre la transmission nativement numérique des procédures entre les services d'enquête et les juridictions. Le déploiement, initialement prévu entre 2020 et 2022, a d'abord ciblé la filière correctionnelle (chaîne pénale des majeurs, délits). L'objectif opérationnel était de commencer par les procédures les plus simples – notamment celles classées sans poursuites pour les infractions mineures (« petits X ») – puis d'étendre progressivement le dispositif aux COPJ et aux déferrements.

La PPN repose sur un bureau pénal numérique (BPN), constitué d'un ensemble d'applications :

- PLINE, pour les échanges internes de pièces dématérialisées au sein du ministère de la Justice ;
- PLEX, destiné à la transmission de pièces volumineuses entre justice et auxiliaires de justice ;
- CEP, permettant les échanges entre juridictions et avocats via e-Barreau ;
- NPP, qui assure la réception des procédures pénales numériques et la vérification de la signature électronique ;
- NOE, outil de préparation des audiences à partir des procédures reçues ;
- NOTIDOC, plateforme d'échanges entre greffiers et huissiers ;
- WiFi Avocat, réseau sécurisé mis à disposition des avocats dans les juridictions.

Mais l'innovation de la PPN ne réside pas seulement dans les outils : elle tient aussi à une méthodologie inédite dans la Justice. Dès 2019, une véritable expérimentation de terrain (qualifiée de « Lab » plutôt que d'expérimentation classique) a été conduite à Amiens et Blois. Contrairement aux expérimentations publiques habituelles, dont l'issue est souvent programmée à l'avance, le Lab a constitué un espace d'itération réelle impliquant simultanément les équipes techniques, les utilisateurs (policiers, gendarmes, greffiers, magistrats) et la direction des affaires criminelles et des grâces (DACG), capable d'adapter en temps réel la norme juridique aux besoins identifiés.

---

[40] Le déploiement à la cour d'appel de Paris n'a pas abouti en 2022.
[41] D. Basso, « Place Vendôme : Deloitte enrôlé pour améliorer Cassiopée, le logiciel des magistrats », La Lettre, 13 octobre 2025

Ce modèle « laboratoire vivant » a permis d'identifier les véritables défis : non pas la signature ou la rédaction numérique, déjà maîtrisées, mais la perte du « hors-procédure » qu'apportait le papier (épaisseur d'un dossier, post-it d'instruction, matérialité de l'urgence). Il a fallu traduire ces indices tangibles en interfaces numériques : codes couleurs, indication de volumétrie, tableaux de réception structurés.

L'expérimentation d'une année a débouché sur la première transmission numérique complète d'une procédure en 2019. Elle a conduit également à une adaptation législative majeure : la loi du 23 mars 2019 a intégré dans le code de procédure pénale le régime juridique du dossier pénal numérique, reconnue comme probant, et permettant les passages papier vers le numérique tout en conservant la valeur des actes signés électroniquement.

Le déploiement national s'est ensuite structuré en vagues successives, en ciblant d'abord les juridictions et services d'enquête présentant la meilleure maturité numérique, afin de garantir des « quick wins » (victoires rapides) indispensables à la crédibilité du programme. Ce déploiement a permis de réaliser le traitement dématérialisé (réception, traitement, archive) des deux tiers des 4,5 millions de procédures annuelles transmises par le ministère de l'Intérieur dans un format numérique probant.

Le programme a néanmoins été confronté à des enjeux techniques structurels : rénovation des outils de rédaction des forces de l'ordre, réflexion sur les capacités de stockage et l'éventuel recours au cloud, ou encore difficultés persistantes liées à la dette technologique du ministère de la Justice.

Enfin, la direction de programme a donné une priorité à la conduite du changement. La culture du papier est profondément ancrée chez les magistrats, greffiers, policiers et gendarmes. Le succès du déploiement a nécessité la mise en place d'un réseau local de chefs de projet PPN, complété par les ambassadeurs de la transformation numérique issus des services judiciaires, capables d'assister les utilisateurs au quotidien, de distinguer les problèmes d'outil des problèmes de réseau, et d'incarner une présence rassurante sur le terrain.

### I.3.3. En matière civile

18. **Le Réseau Privé Virtuel Avocat (RPVA)** – Alors que s'était déployé le réseau intranet dans les services du ministère de la justice à la fin des années 1990 (RPVJ, Réseau Privé Virtuel Justice), le Réseau Privé Virtuel Avocat (RPVA) est mis en œuvre au milieu des années 2000 et il est encore aujourd'hui le support de communication électronique pour la profession.

Il doit être rappelé que le premier périmètre de la convention du 4 mai 2005 entre les avocats et le ministère de la justice était double : la gestion prévisionnelle des référés et la mise en état électronique. Il était alors prévu de transmettre par voie électronique uniquement les actes de procédure et des messages avec une gestion des affaires, sans les pièces du dossier de fond (la limite des serveurs était fixée à 4 Mo au départ).

Mais les avocats y ont rapidement vu une commodité pour transmettre des pièces numérisées de leur dossier, si bien que beaucoup de critiques sur les limites de l'outil viennent en réalité du détournement de son objectif pour transmettre à la fois des actes et des pièces. Au vu du besoin des utilisateurs, la capacité de transmission a été ensuite accrue jusqu'à 10 Mo.

19. **La dématérialisation de la procédure civile** – Pour mettre en œuvre concrètement la dématérialisation de la procédure, a donc été développé, à partir de la chaîne civile informatique WinCI, un module de communication électronique dénommé « ComCi » permettant l'échange de manière sécurisée d'un certain nombre de données et de documents, entre les tribunaux judiciaires et les cabinets d'avocats, notamment afin de permettre une mise en état « virtuelle ».

Ce module communique avec le système informatique des cabinets d'avocats par le biais d'une interconnexion entre le RPVA et le RPVJ des tribunaux pour garantir la sécurité des échanges. Il devrait également être cité la communication avec les experts (OPALEXE), les huissiers de justice[42] (Réseau Privé Sécurisé des Huissiers de Justice) et la gestion des injonctions de payer (IPWEB) qui constituent les différentes briques de la communication électronique civile. Ces réalisations ont pour particularité d'avoir été développées sous l'impulsion des professions concernées, avec l'appui des services de l'administration centrale, dans un souci d'intérêt partagé.

20. **Les premiers outils de la gestion des affaires civiles** – Au titre des logiciels cités l'histoire de « WinCi » pourrait être approfondie pour son originalité. Issu du début des années 1990 dans un développement réalisé avec un atelier de génie logiciel « WinDev » par la société AROBASE (devenue ESABORA), le logiciel a émergé durant l'âge d'or de l'initiative locale et a connu de multiples déclinaisons en France et dans le monde, même au-delà du secteur de la justice pour assurer, la gestion de processus (« affaires »)[43]. Non communicant à l'origine (un serveur local par tribunal), il s'est appuyé ensuite sur le RPVJ, avec l'adjonction du module « ComCi » pour permettre la communication électronique. Cette suite logicielle est toujours fonctionnelle en 2025.

Le projet Transjuris, préfigurateur du lancement de Portalis (avec une date initiale de démarrage des travaux en fin 2000), devrait aussi être cité. Créé par le ministère de la justice et la Caisse des dépôts et consignations en 2008 pour une durée de quatre ans, le projet a eu pour ambition de réunir tous les outils existants et d'en construire un nouveau avec les professions du droit et de la sécurité publique. Le projet a été stoppé après 2012.

---

[42] Devenus Commissaire de justice depuis le 1er juillet 2022
[43] V. les références de la société ESABORA : https://esabora.com/nos-references/

21. **La centralisation des outils de gestion des affaires civiles : le projet Portalis** – Dans les années 2010, c'est au tour du projet « Portalis » de prendre corps pour chercher à remplacer les applications de gestion des affaires civiles dans les cours et tribunaux, dont le cœur a été conçu dans les années 1990 (logiciels de la série WINCI). Après s'être opportunément adressé aux usagers dans le mitan des années 2010 avec le portail informatif du justiciable puis le portail applicatif du justiciable et celui du Service d'accueil unique du justiciable (SAUJ[44]), afin d'éviter un « effet tunnel » dû à la complexité d'élaborer les solutions de « back office ». Le projet s'inscrit toujours dans une dynamique de développement au début des années 2020 avec le portail des auxiliaires de justice et le bureau virtuel métiers. Un nouvel applicatif métier et une dématérialisation totale de la chaîne civile sont également inclus dans ce projet.

Même si Portalis devait être réalisé en fin 2010, sous l'impulsion de Transjuris, sa réalisation effective ne débute en réalité qu'au début des années 2010. Le 7 juillet 2014, le ministère de la justice décide de concentrer ses moyens pour produire une offre de services aux justiciables non représentés et de geler l'évolution des outils préexistants construits avec les professions du droit (ComCi, OPALEXE, RPSH et IPWEB). Un premier décret du 11 mars 2015, a permis d'expérimenter les échanges par messagerie électronique avec les justiciables non représentés mais s'est révélé chronophage pour le greffe et peu sécurisé (simples messages électroniques). Cet arbitrage a été remis en cause en février 2022, à la suite du rapport d'étape de la Cour des Comptes de janvier 2022, pour orienter à nouveau le projet vers les professionnels.

Au moment de la clôture de cette étude, l'actualité du projet Portalis est marquée par une importante étape juridique avec l'arrêté du 31 janvier 2025 qui autorise officiellement la mise en œuvre d'un traitement automatisé de données à caractère personnel dénommé « Portalis – Portail des juridictions », couvrant l'ensemble des procédures civiles et sociales devant les tribunaux judiciaires, conseils de prud'hommes, tribunaux paritaires des baux ruraux et cours d'appel. Portalis est toujours présenté comme l'outil de gestion et de suivi unique que toutes les juridictions civiles doivent progressivement adopter d'ici 2027, dans la continuité des déploiements déjà réalisés pour le portail du justiciable et la communication électronique avec les avocats. Portalis figure parmi les grands projets informatiques prioritaires, bénéficiant de crédits dédiés pour poursuivre les développements et les déploiements, tandis que les échanges avec le Conseil national des barreaux ont permis de stabiliser un scénario de mise en œuvre qui prévoit une coexistence transitoire avec les applications existantes avant une bascule complète vers la nouvelle chaîne numérique.

22. **Les opportunités de la juridiction unique des injonctions de payer** – Des projets de réforme, comme l'unification du contentieux des injonctions de payer au sein d'une juridiction nationale (JUNIP), auraient pu aussi employer de nouveaux moyens algorithmiques pour faciliter le traitement de ce contentieux technique. L'abandon du projet, pour diverses raisons comme le souhait de rapprocher les justiciables de leurs juridictions, n'a toutefois pas permis cette démonstration[45].

---

[44] Art. R123-28 du code de l'organisation judiciaire

### I.4. Depuis le milieu des années 2010 : le foisonnement d'une offre privée innovante, poussant l'administration de la justice à accélérer son adoption de nouvelles technologies

#### I.4.1. Un foisonnement d'offre privée innovante

23. **La poussée de legaltechs au cœur de l'innovation** – Depuis le milieu des années 2010, une troisième vague de transformation numérique est en marche, mais se joue principalement en dehors des tribunaux. Cette vague se caractérise notamment par l'émergence d'une multiplicité d'offres de services par le secteur privé, s'appuyant sur des technologies de rupture telles que « l'IA » ou les *blockchain*s, avec l'ambition d'apporter cette fois-ci une valeur ajoutée sur l'interprétation du droit et sa mise en œuvre, en plus de dématérialiser pleinement les échanges.

    L'originalité de cette période est l'entrée sur le marché d'entrepreneurs n'ayant que peu ou pas de lien avec les professions du droit, en rupture avec la première période étudiée, où l'initiative était étatique, et la deuxième, avec initiative partagée avec les professionnels[46]. Cette troisième vague pousse toutefois le secteur public à mieux s'approprier les dernières technologies et a conduit le ministère de la justice à diversifier ses approches.

24. **Origine de la poussée des legaltechs** – Les premières legaltechs sont apparues aux États-Unis dans les années 2000 dans le contexte de la crise financière en réponse aux problèmes rencontrés par les métiers du droit comme la charge croissante de régulation imposées aux entreprises et la recherche de réduction du coût des procédures juridiques.

    En 2010, le projet d'analyse juridique Lex Machina sort du cadre de l'université de Stanford pour devenir une startup, pionnière de l'analyse « prédictive » en droit des affaires. En parallèle, la justice et la police commencent à tester des solutions s'appuyant sur l'IA : dès 2011, la police de Los Angeles adopte l'algorithme PredPol pour anticiper les lieux de crimes à partir des données historiques.

    La justice civile américaine, qui admettait déjà l'emploi d'algorithmes de découverte de preuves (electronic discovery) depuis des amendements aux règles civiles de procédures depuis 2006, va également admettre la possibilité d'employer l'apprentissage automatique pour l'exploitation de documents (technology-assisted review, TAR) par les parties (Southern District of New York, Rio Tinto vs Vale, 2015). Des recherches académiques explorent aussi le potentiel de l'IA dite prédictive : par exemple, des modèles statistiques « prédisent » en 2014 les décisions de la Cour

---

[45] « Injonction de payer dématérialisée : suppression de la juridiction unique (JUNIP) », *Éditions législatives*, 20 mai 2021, accessible sur : https://www.editions-legislatives.fr/actualite/injonction-de-payer-dematerialisee%C2%A0-suppression-de-la-juridiction-unique-junip/, consulté le 3 mars 2022

[46] Cette tendance est également à constater dans des domaines comme la santé.

suprême des États-Unis avec environ 70 % de précision, ouvrant la voie à un grand engouement pour la « justice prédictive »[47].

En 2015, LexisNexis rachète Lex Machina, marquant la première acquisition d'une startup d'analytique juridique basée sur l'apprentissage automatique. En 2016, la société ROSS Intelligence, fondée en 2014, dévoile ROSS, présenté comme la première IA capable de répondre, comme un véritable avocat, aux questions juridiques en langage naturel grâce à IBM Watson[48]. D'autres entreprises émergent également à cette époque comme Kira Systems, LawGeex et Ravel Law.

La vague est telle qu'en 2023, une note d'analyse de Goldman Sachs estimait que 44% des emplois dans le domaine juridique étaient désormais exposés à un remplacement par « l'IA »[49].

En Europe, Luminance est créé au Royaume-Uni en 2016 et automatise la recherche juridique et la révision de contrats. La même année, une étude de l'University College London (UCL) est fortement médiatisée : un modèle entraîné sur 584 affaires parvient à classer correctement les arrêts de la Cour européenne des droits de l'homme, entre violation et non violation de la Convention, dans 79 % des cas[50].

25. **La situation en France** – En France, les premières legaltechs exploitant « l'IA » apparaissent également vers 2016 : Predictice et Doctrine proposent des moteurs de recherche et d'analyse de jurisprudence alimentés par l'apprentissage automatique. C'est la promesse de pouvoir anticiper, pour certains contentieux civils ou commerciaux, les fourchettes de montants pouvant être prononcés par les tribunaux qui attire alors l'attention sous le vocable contesté de « justice prédictive »[51]. Caselaw Analytics, se présentant comme évaluant le risque judiciaire par diverses méthodes algorithmiques dont de l'apprentissage automatique, est créé en 2017. L'accueil par les professionnels est contrasté, certains magistrats n'y trouvant qu'une faible valeur ajoutée[52] alors que d'autres y voyaient une collégialité élargie[53].

26. Les éditeurs juridiques historiques enrichissent leurs offres de moteurs de recherche avec des outils d'analyse. Depuis fin 2022, c'est au tour des « IA » de génération de contenu, du type ChatGPT, d'investir le terrain médiatique en s'imposant comme un complément des moteurs de recherche pour générer des analyses à destination de professions comme les avocats et les juges[54]. Les éditeurs ont donc complété leur offre avec de tels outils génératifs (GenIA-L chez Dalloz, LexisNexis+IA chez LexisNexis), tout comme les legaltechs (Ordalie par exemple)[55].

27. **Le développement des blockchains** – La *blockchain*, qui est une technologie de stockage et de transmission d'informations sans organes de contrôle connue notamment pour supporter les cryptomonnaies, a été non seulement employée pour servir de nouveaux tiers de confiance afin d'enregistrer des actes, mais également pour suivre l'exécution automatisée de contrats (*smart contracts*)[56]. Ces nouvelles formes de contrat ont pour particularité de ne plus contenir de clauses écrites mais des règles s'exécutant de manière automatisée.

### I.4.2. Une inspiration pour le secteur public, avec des nouvelles technologies peinant encore à se banaliser dans les tribunaux

28. **Des technologies de rupture peinant à se banaliser dans les tribunaux** – Du côté du secteur public, ces technologies de rupture n'ont pas encore pleinement pénétré les usages, même si l'emploi d'outils du secteur privé de recherche de jurisprudence ou l'emploi « d'IA » générative hors cadre conquiert un nombre croissant d'utilisateurs.

Les 3 éditions des « Vendôme Tech » entre 2017 et 2019, ainsi que les « Rendez-vous de la transformation du droit », ont été l'occasion pour les services du ministère de prendre acte des opportunités et lancer des initiatives, mais l'emploi au quotidien de technologies avancées dans les juridictions demeure pour l'essentiel le fait d'initiatives personnelles[57].

Ainsi, le projet Datajust, qui visait à employer de l'apprentissage automatique pour établir des échelles d'indemnisation en matière de réparation du préjudice corporel, n'a pas dépassé le stade de l'expérimentation en raison de sa complexité technique excessive[58]. Parmi les éléments caractérisant cette complexité, il peut être cité l'objectivation de la pertinence de 40 critères pour caractériser le traitement de la réparation du préjudice corporel, la nécessité de moyens importants pour prévenir les biais algorithmiques, et une base de données incomplète, limitée aux décisions d'appel sans inclure les premières instances, ce qui générait des biais. De plus, le projet a suscité des critiques sur la protection des données personnelles et l'absence de

---

concertation avec les professionnels du droit. L'abandon a été acté le 13 janvier 2022, à la fin de la période expérimentale de deux ans autorisée par le décret du 27 mars 2020, sans prolongation possible.

Au niveau local, des initiatives ont conduit à des développements très pratiques, comme la dématérialisation de certains processus de travail d'un parquet[59] ou le traitement du contentieux de la consommation.

29. **Contexte de l'apparition de l'application JCP** – L'exemple de l'application éditique JCP, développée par un juge des contentieux de la protection (JCP) pour ses pairs, constitue un cas particulièrement révélateur de l'appropriation de technologies hors des processus officiels de conception. Conçue pour alléger le travail de rédaction en matière de contentieux de masse, cette application web a eu pour ambition d'automatiser des opérations répétitives telles que le remplissage de trames. L'objectif du concepteur a été de se libérer du temps pour se concentrer sur la rédaction de motivations au fond des jugements complexes.

30. **Au-delà de la fonction éditique** – L'outil a ensuite évolué au-delà de sa vocation éditique première, en intégrant un module automatisé de vérification des contrats de crédit à la consommation. Celui-ci permet le calcul du taux annuel effectif global, la comparaison avec le seuil de l'usure (seuil qui varie trimestriellement) et facilite, le cas échéant, la transmission automatique au parquet d'infractions constatées. Cette fonctionnalité répond à une obligation dont l'exercice systématique se révélait difficile en raison de la technicité des calculs et de la charge de travail des magistrats. L'application a été massivement adoptée par les juges concernés, atteignant plusieurs dizaines puis plus d'une centaine d'utilisateurs réguliers. Elle a également suscité un intérêt au-delà de la sphère judiciaire, notamment de la part des autorités chargées du contrôle administratif des pratiques bancaires, qui y ont trouvé un outil susceptible d'améliorer leurs propres opérations de vérification.

31. **Les limites d'une solution d'initiative locale** – Cette diffusion progressive a toutefois mis en lumière les limites d'un développement hors cadre institutionnel. Les interrogations relatives à la sécurité, à la responsabilité, à la propriété intellectuelle ou encore à la transparence du code ont révélé l'ambivalence des initiatives locales : utiles, efficaces, parfois indispensables, elles demeurent en tension avec les règles propres à un système d'information public. L'outil a ainsi incarné la figure emblématique d'un développement de « Shadow IT » en matière judiciaire, révélatrice à la fois d'un besoin non satisfait et d'une inventivité bienvenue sous un angle opérationnel, mais fragile par son manque d'ancrage institutionnel.

---

[59] Les travaux dans un parquet de l'Est de la France, encore en l'état de projet au moment de cette étude, ne seront pas développés ici. Il a été privilégié l'application JCP déjà en production.

**II.** **La portée de la transformation numérique de la justice : les ambitions à l'épreuve de la réalité**

32. **Plan** – Les ambitions et annonces s'étant succédées de manière très volontariste au ministère de la justice ont toutefois eu bien du mal à se concrétiser dans le quotidien des juridictions.

Les évaluations réalisées par les instances européennes confirment les efforts, avec des résultats contrastés (2.1). Au niveau national, l'insuffisance de l'évaluation des besoins et de la mesure du retour sur investissement ne permet pas de piloter efficacement la conduite du changement (2.2). Dans ce contexte, les critiques objectives et plus subjectives s'entremêlent, entre critique classique de la nouvelle gestion publique (2.3), craintes de la déshumanisation et du remplacement des professions (2.4) et, vu du terrain, l'insuffisante capitalisation des besoins des utilisateurs en juridiction (2.5).

**II.1.** **L'évaluation de la transformation numérique de la justice en France**

33. **Plan** – Il ne sera pas réalisé ici une description des méthodologies d'évaluation des instances européennes, ni leur critique, les « scores » attribués étant présentés de manière brute et simplement comparés aux autres États évalués. Les deux documents étudiés (tableau de bord justice de la Commission européenne – 2.1.1 et rapport de la CEPEJ – 2.1.2) convergent pour situer la France dans le dernier tiers de leurs évaluations.

**II.1.1.** *L'évaluation de l'Union européenne (tableau de bord justice de la Commission européenne)*

34. **Présentation du tableau de bord** – Le tableau de bord de la justice de l'Union européenne (EU Justice Scoreboard) est un outil annuel de la Commission européenne qui compare le fonctionnement des systèmes judiciaires des États membres. Il fournit des données objectives sur l'efficacité, la qualité et l'indépendance de la justice, afin de soutenir les réformes nationales, de favoriser l'État de droit et d'améliorer le climat d'investissement et de confiance dans l'UE.

35. **Évaluation du niveau d'utilisation des technologies numériques dans les tribunaux** – Dans ce cadre, la Commission européenne intègre la dimension numérique en attribuant un score de 0 à 7 pour évaluer le niveau d'utilisation de différentes technologies par les tribunaux et les services de poursuite.

La France se situe dans le dernier tiers des États membres avec un score de 3,5 pour les tribunaux et de 3 pour les parquets. Si l'emploi de la visioconférence apparaît comme banalisé et les systèmes de gestion d'affaires en matière pénale, tous les autres secteurs (blockchains, IA, travail à distance, distribution automatique d'affaires) paraissent en retrait. L'Estonie et l'Allemagne se situent en haut du premier tiers des États évalués (avec un score de 6), ainsi que l'Autriche (4,5 pour les tribunaux et 6 pour les parquets).

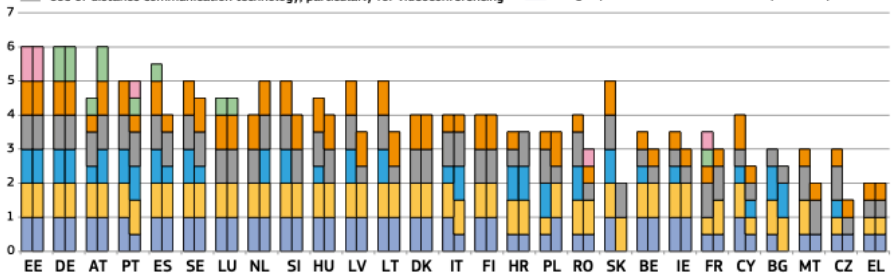

Source : The 2024 EU Justice Scoreboard (données 2023)

### II.1.2. L'évaluation du Conseil de l'Europe (rapport de la Commission européenne pour l'efficacité de la justice – CEPEJ)

**36. Présentation du rapport de la CEPEJ** – Le rapport d'évaluation de la CEPEJ[60] analyse de manière détaillée et comparative le fonctionnement des systèmes judiciaires de plus des 46 États membres du Conseil, sur la base de données approfondies collectées auprès des institutions nationales. Contrairement au tableau de bord justice de la Commission européenne, qui couvre uniquement les États membres de l'UE et met l'accent sur quelques indicateurs clés pour éclairer les politiques européennes et le Semestre européen, le rapport de la CEPEJ propose une analyse bien plus exhaustive, méthodologiquement détaillée et à vocation comparative paneuropéenne, sans finalité de pilotage direct des politiques économiques de l'UE. Le rapport ambitionne en revanche de participer à prévenir notamment les violations de l'art.6 de la Convention européenne des droits de l'homme (procès équitable, dont les questions relatives au respect d'un délai raisonnable).

**37. Évaluation du niveau d'utilisation des technologies numériques dans les tribunaux** – Dans ce cadre, la CEPEJ évalue également l'utilisation des technologies de l'information et de la communication (TIC) en attribuant un score de 0 à 10, combinant trois composantes : gestion des affaires, accès numérique à la justice et aide à la décision.

Avec un score d'utilisation de 2,88 sur 10, la France se situe en dessous de la médiane des États évalués (3,4). Une singularité est à relever : le fort écart entre le score entre la justice civile (1,8) / pénale (1,4) et la justice administrative (5,3). L'accès numérique est globalement évalué à 1,7, l'aide à la décision à 2,2 et la gestion des affaires à 4,5. L'Estonie dispose du meilleur score selon l'évaluation de la CEPEJ (7,6), suivi de l'Espagne (6,1) et de l'Italie (6,05).

---

[60] Commission européenne pour l'efficacité de la justice, corps dépendant du Conseil de l'Europe et regroupant des représentants désignés par les ministères de la Justice des 46 États membres de cette organisation internationale.

Carte 6.7 **Indice d'utilisation des TIC**

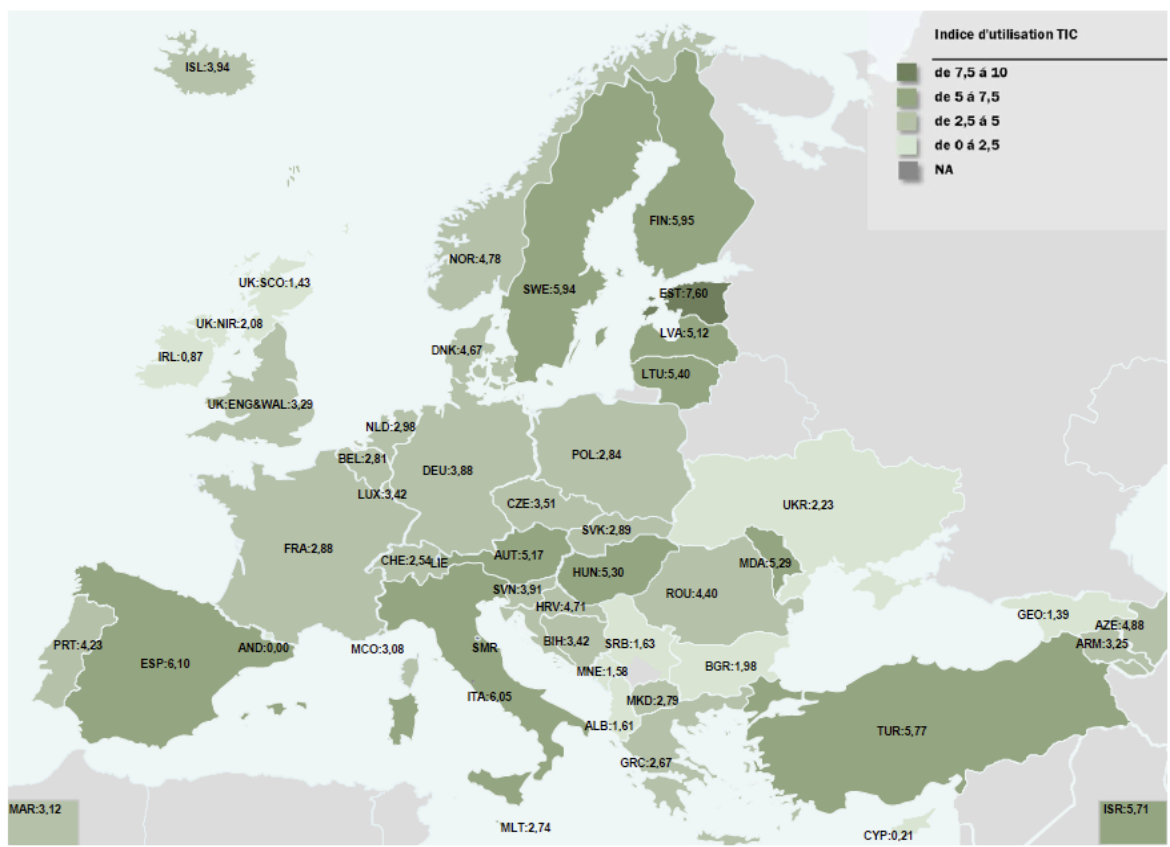

Source : Systèmes judiciaires européens, rapport d'évaluation de la CEPEJ, cycle d'évaluation 2024 (données 2022)

## II.2.    Un retour sur investissement encore mal mesuré

38. **La difficile mesure du lien entre informatisation et performance** – L'objectivation de ces évaluations européennes est rendue complexe du fait de l'absence de mesure précise des effets concrets de ces technologies sur la justice.

La CEPEJ a tenté elle-même, en 2016, de rechercher une éventuelle corrélation entre taux d'informatisation et performances des tribunaux[61]. À partir des données collectées, elle conclut que c'est probablement moins la multiplication de services numériques que la manière de conduire des projets informatiques qui semblerait produire des effets[62]. Penser l'organisation d'un processus de travail et l'appuyer d'un outil informatique pour l'exécuter semblerait ainsi plus efficace que de déployer un outil informatique et de rechercher ensuite l'organisation optimale.

---

[61] Systèmes judiciaires européens, efficacité et qualité de la justice – Rapport thématique : l'utilisation des technologies de l'information dans les tribunaux en Europe », *op.cit.*
[62] V. « Lignes directrices sur la conduite du changement vers la Cyberjustice », CEPEJ, 2017

Des rapports en France sont parvenus à un constat relativement similaire, en incitant les responsables publics à mieux prendre en compte des attentes des utilisateurs et à mesurer précisément le retour sur investissement[63].

39. **Ne pas concevoir le numérique comme une finalité** – De même, la mesure de l'impact des récentes innovations venant du secteur privé n'est pas aisée. Guy Canivet, premier président honoraire de la Cour de cassation et ancien membre du Conseil Constitutionnel, préconisait dans le rapport *Justice : faites entrer le numérique[64]* de ne pas tirer de conclusions hâtives au sujet de ces outils et de revenir aux attentes des justiciables et d'en faire le point de départ pour toute transformation numérique. Il faut dire que les attentes des justiciables ne portent pas prioritairement tant sur un accès en ligne à la justice que sur une réponse efficace.

Ainsi le SIAJ (système d'information d'aide juridictionnelle) est uniquement ouvert aux justiciables, pour certains tribunaux judiciaires (en expérimentation et en excluant les tribunaux administratifs, des formulaires CERFA étant encore à utiliser).

Pour ses détracteurs, cette offre de service a été ouverte en méconnaissant la réalité de la vie d'un cabinet et du début de l'accompagnement d'un processus d'un contentieux. Les justiciables ne commencent pas toujours par solliciter l'aide juridictionnelle, mais par obtenir les conseils d'un avocat qu'ils auront choisi pour sa compétence dans leur contentieux. De plus, l'avocat, qui n'aura pas accès au suivi en ligne, n'incitera probablement pas ses clients à employer le service. Il aurait semblé plus intéressant d'envisager aussi l'hypothèse d'une prise en charge intégrale de l'action au fond des justiciables, y compris de leurs demandes d'aide juridique qui est l'accessoire de leurs démarches.

## II.3. L'informatique saisie par les critiques de la nouvelle gestion publique

40. **Des discours absorbés par une critique politique** – Une autre catégorie de critiques, appropriées notamment par les syndicats de personnels, est relative à ce qui est qualifié par certains d'emprise managériale croissante sur l'organisation de la justice[65] qui aurait pris corps au sein de plusieurs courants de réforme[66].

Soulignons que si le besoin de réformer la justice et la nécessité d'une bonne gestion des fonds publics font plutôt consensus, ce sont plutôt les faiblesses des études d'impact des différentes réformes (qui minimiseraient les incidences sur l'informatique) et le manque d'analyse du travail réellement fourni par les tribunaux pour s'adapter à celles-ci qui sont caractérisées par certains détracteurs[67]. Pour eux, le numérique n'a pas forcément fourni de solution pour traiter l'augmentation des contentieux de masse[68], ce qui expliquerait pourquoi l'initiative publique se trouve débordée ces dernières années par le dynamisme des initiatives privées. Un dynamisme d'ailleurs moins contesté pour sa capacité à apporter de nouvelles solutions que pour le projet de justice qu'il sous-tend.

---

[63] V. le rapport de la mission d'information d'Etienne Blanc relatif aux carences de l'application des peines et à l'évaluation de l'application Cassiopée – Rapport n°3177 déposé à la présidence de l'Assemblée nationale le 16 février 2011 ou le rapport conjoint Inspection des finances / Conseil général de l'économie, de l'industrie, de l'énergie et des technologies, le pilotage et l'audit des grands projets informatiques de l'État, mars 2012
[64] Institut Montaigne, Justice : Faites entrer le numérique, *op.cit.*
[65] E. Serverin, « Comment l'esprit du management est venu à l'administration de la justice » in B. Frydman (dir.), E. Jeuland (dir.), *Le nouveau management de la justice et l'indépendance des juges*, Dalloz, 2011, p.54
[66] A. Vauchez, Laurent Willemez, *La Justice face à ses réformateurs*, PUF, 2007, p.118
[67] E. Poinas, « Comment les nouvelles technologies ambitionnent de révolutionner la fonction de juger », *Actu-Juridique.fr*, Lextenso, 15 avril 2019

Les solutions numériques ont pu aussi servir d'outil à des réformes, comme la réforme de la carte judiciaire en 2008. Ainsi, en matière pénale, aux difficultés techniques de CASSIOPEE, couplées aux réorganisations imposées, s'est ajouté le reproche de servir d'instrument à une réforme décriée par les divers syndicats.

En matière civile, la communication électronique avec les avocats, limitée aux référés et aux procédures civiles ordinaires, a conduit à des critiques de remplacement des juridictions physiques par des juridictions virtuelles afin de réduire les coûts.

### II.4.    Les craintes de déshumanisation et le remplacement des professions

41. **Des discours à situer dans la critique de la technique** – Parmi les autres craintes habituelles souvent relayées s'agissant de la transformation numérique de la justice, l'on retrouve également un ensemble de reproches sur la mécanisation de l'institution et les risques de remplacement par des algorithmes. Ces discours, typiques de la critique de la technique, se retrouvent massivement dans la plupart des professions où la sophistication des dernières générations d'algorithmes permet d'envisager des traitements exigeant auparavant une expertise humaine.

Le recours à « l'IA » a ravivé des craintes assez anciennes d'automatisation et de déshumanisation, alimentées par un traitement relativement imprécis du phénomène par les médias[69].

Ainsi, l'analyse statistique de la jurisprudence avec de nouveaux moyens (jurimétrie ou justice dite prédictive, prévisionnelle, actuarielle, quantitative) a cristallisé un grand nombre de critiques en laissant planer la crainte d'un possible remplacement du juge[70] alors qu'elle intéresse en tout premier lieu certaines professions, comme les avocats, les assureurs ou les directions juridiques d'entreprises, non pas chargées de « dire le droit », mais d'évaluer des risques[71].

Les *blockchains* ont aussi conduit des commentateurs à présumer l'obsolescence de la profession de notaire alors que celle-ci s'en est plutôt bien emparée pour tenter de résoudre le problème de durée de vie des certificats de signature électronique[72].

**II.5. La qualité de la capitalisation des besoins des utilisateurs contestée par le terrain**

*II.5.1. Le constat d'une inadéquation entre les applications nationales et les besoins du terrain*

42. **Un écart entre besoins et solutions persistant** – Depuis plusieurs années, les juridictions soulignent un écart persistant entre les besoins opérationnels, bien perçus des agents, et le contenu des solutions numériques développées au niveau central.

Cet écart se manifeste particulièrement dans le champ de l'éditique, qui constitue pourtant l'un des instruments essentiels du travail juridictionnel. La rédaction des décisions, l'adaptation des trames aux réformes successives et la fusion des données procédurales représentent une part significative de l'activité quotidienne des magistrats comme des greffiers. Toutefois, les outils fournis ne répondent qu'imparfaitement à ces exigences, que ce soit en raison de leur architecture ancienne, de difficultés de maintenance ou du délai nécessaire pour intégrer les nombreuses évolutions législatives et réglementaires.

43. **Les causes de cet écart entre les besoins et les solutions** – Il y a plusieurs causes de cette inadéquation. D'une part, la fréquence des réformes législatives entraîne des besoins soutenus d'actualisation des trames, que les applicatifs centraux peinent à absorber dans un temps compatible avec l'activité des juridictions. D'autre part, le modèle d'uniformisation des trames, élaborées au niveau national puis déclinées localement, se heurte à la très forte « indépendance rédactionnelle » des magistrats, qui manifestent une forte méfiance et beaucoup de réticence à l'égard de modèles perçus comme trop standardisés. Et lorsque des processus de construction collective de trames ont été expérimentés, les consensus entre rédacteurs sont souvent difficiles à atteindre du fait de pratiques hétérogènes, n'étant pas nécessairement conciliables.

À cela s'ajoutent des contraintes matérielles, comme l'usage d'un traitement de texte propriétaire ancien (WordPerfect), limitant l'appropriation et contraignant les tentatives locales de modernisation.

44. **La réaction des acteurs du terrain à cette inadéquation** – Dans ce contexte, l'initiative locale reste vivace, entre développements spécifiques ponctuels[73] et usages non déclarés de solutions commerciales pour des tâches professionnelles (« Shadow IT » et « Shadow AI »), avec des risques majeurs de fuite de données confidentielles. Ces phénomènes, dont l'ampleur est mal mesurée en France (selon une étude de l'UNESCO dans 96 pays, 44% des professionnels de la justice utiliseraient des outils d'IA générative[74]), soulèvent des enjeux déontologiques, de transparence et de sécurité majeurs. La Cour constitutionnelle de Colombie a établi des limites strictes de l'utilisation de « l'IA » par des juges à la suite d'une décision où un juge d'appel avait

---

[73] *Cf. supra* avec l'exemple de l'application pour les juges des contentieux de la protection
[74] UNESCO Global Judges' Initiative: survey on the use of AI systems by judicial operators, CI/DIT/2024/JI/01 Rev, 2024, accessible sur : https://unesdoc.unesco.org/ark:/48223/pf0000389786, consulté le 30 novembre 2025

interrogé ChatGPT sur une question de droit et retranscrit sa démarche dans la motivation de l'arrêt[75].

### II.5.2. L'évolution de la doctrine ministérielle : d'une méfiance de l'initiative locale à l'accompagnement

45. **L'initiative locale : de la méfiance à l'accompagnement** – Après les années de foisonnement de l'initiative locale entre 1980 et 2000, le ministère de la justice a encadré plus strictement les initiatives émanant des juridictions, accueillant avec beaucoup de rigueur, voire de méfiance, les initiatives applicatives pouvant émerger[76]. Mais, durant ces dernières années, le ministère de la justice a fait évoluer de manière notable sa doctrine, en s'appuyant sur des dispositifs interministériels.

46. **Un accompagnement employant des dispositifs interministériels** – Ainsi, le développement du dispositif des startups d'État, au sein de l'incubateur ministériel, constitue l'une des inflexions les plus importantes. Inspiré du manifeste beta.gouv, ce dispositif place explicitement les besoins des utilisateurs avant ceux de l'administration, privilégie l'expérimentation rapide à partir d'un problème bien identifié et confère aux équipes une grande autonomie dans les choix technologiques comme dans la conduite du projet.

Les startups d'État sont conçues comme des structures souples, dotées de développeurs, designers et chefs de produit, pouvant rencontrer les usagers, tester des prototypes en quelques jours, mesurer leur impact en situation réelle et réorienter rapidement les développements. Ce mode opératoire contraste avec les projets informatiques plus classiques, marqués par des cahiers des charges volumineux, des cycles longs et une faible interaction avec les agents de terrain.

47. **Un accompagnement avec des dispositifs spécifiques au ministère de la justice** – Des programmes complémentaires, comme les prototypes (POC, *proof of concepts*) portés par l'Atelier Numérique pour la Justice (ANJe) repris d'initiatives locales, témoignent également d'une volonté de créer des passerelles entre les innovations issues des juridictions et les outils nationaux.

48. En permettant d'évaluer la pertinence, la faisabilité et l'impact d'une solution conçue localement, ces dispositifs offrent une voie légitime de reconnaissance et, le cas échéant, de généralisation. Ils visent à éviter que des outils efficaces ne restent confinés à des pratiques informelles, et à capitaliser sur l'ingéniosité des agents sans renoncer aux exigences de sécurité et de cohérence du système d'information ministériel.

---

[75] M. Fandino, J. del Carmen Ortega, S. Gil, « Comment un juge devrait-il utiliser l'IA ? Une première réponse avec une décision de la Cour constitutionnelle de Colombie, Village de la Justice, 19 septembre 2024, accessible sur : https://www.village-justice.com/articles/comment-juge-devrait-utiliser-une-premiere-reponse-une-cour-constitutionnelle,50871.html, consulté le 29 novembre 2025

[76] Parmi les exemples d'initiative locale capitalisé par le ministère de la justice pour les tribunaux, il peut être cité par exemple l'outil PILOT, développé au début des années 2000 pour planifier les audiences, gérer les absences et piloter les juridictions par un agent en poste à Metz. V. rapport des États généraux de la justice, Le numérique pour la Justice, p.23, accessible sur : https://www.justice.gouv.fr/sites/default/files/2023-11/numerique_egj.pdf, consulté le 15 novembre 2025

### II.5.3. Une dynamique d'accompagnement volontaire

49. **Des besoins émanant des juridictions mieux capitalisés** – D'un côté, les juridictions démontrent une capacité significative à identifier des besoins précis et à construire même parfois des solutions opérationnelles, souvent plus proches de leur réalité quotidienne que celles issues de développements nationaux. De l'autre, l'administration centrale s'est progressivement dotée de mécanismes destinés à mieux intégrer ces initiatives, à sécuriser les développements et à favoriser la diffusion des innovations pertinentes. L'essor des startups d'État et des dispositifs associés illustre le succès de ces approches pour mieux structurer le potentiel issu du terrain.

50. **Une nouvelle dynamique ne couvrant pas encore tous les besoins** – L'existence d'une application comme l'outil éditique JCP, devenu un instrument partagé par de nombreux magistrats, illustre le potentiel des initiatives issues du terrain encore non capté par les dispositifs officiels. S'agissant de cet outil éditique, la raison principale est à trouver dans le souhait du concepteur de continuer à exercer à titre principal son office de juge et non pas de devenir, même pour un temps, un développeur informatique. L'administration n'ayant pas la capacité de s'approprier de tels développements personnels sur ses propres moyens, l'initiative est restée hors cadre.

51. **Une dynamique ayant vocation à se poursuivre** – L'enjeu pour les années à venir consistera à continuer de concilier ces deux dynamiques, en permettant aux innovations locales de bénéficier d'un cadre qui n'en altère ni la réactivité ni la pertinence, tout en renforçant la capacité de l'administration centrale à concevoir ou à intégrer des outils réellement adaptés aux usages professionnels.

La capitalisation des besoins ne pourra être pleinement effective qu'à cette condition : reconnaître le terrain comme producteur de solutions et non seulement comme demandeur (ou un « client »), et structurer un écosystème où l'initiative locale et l'expertise centrale se renforcent mutuellement.

III. **Les perspectives de la transformation numérique de la justice : recomposition de l'offre, ambitions renouvelées, valorisation de la donnée et convergence des acteurs**

III.1. **Vers une recomposition globale et profonde de l'offre de justice**

52. **Vers une privatisation de la justice ?** – La troisième vague de transformation numérique, à l'œuvre depuis le milieu des années 2010 et d'initiative privée, n'est pas qu'instrumentale et interroge de manière profonde l'offre traditionnelle de justice. Si le modèle économique de cette justice du XXIe siècle peine encore à se définir en 2025, la remise en cause de son modèle classique s'affirme sous le double effet d'une mise en concurrence de l'offre publique par une offre privée (parfois de plus faible coût) et d'une désintermédiation.

Ainsi les règlements alternatifs des litiges, encouragés par la loi de programmation et de réforme de 2019, peuvent désormais trouver un support électronique pour leur activité au travers de plateformes en ligne certifiées mettant directement en relation les parties avec un tiers au litige[77]. En France, l'une des premières offre, Médicys, a été créée à l'initiative de la Chambre nationale des huissiers de justice pour encourager la médiation[78]. Les outils d'analyse statistique de la jurisprudence dits de jurimétrie, déjà cités, intègrent même certaines de ces plateformes afin de proposer ce qui pourrait être le probable montant d'indemnisation prononcé par les tribunaux et ainsi définir la meilleure stratégie judiciaire[79].

Dans ce contexte, les legaltechs viennent non seulement soutenir des politiques publiques de réforme de la justice, visant à réduire le nombre d'affaires dans les tribunaux en les orientant vers d'autres modes de traitement, mais créent également de nouveaux segments de service juridique adaptés à des services nativement numériques comme les *smart contracts*.

53. **Le risque d'une fracture entre professionnels** – Au-delà de la seule question des nouveaux acteurs, se pose aussi celle des inégalités d'accès aux technologies avancées comme les « IA » spécialisées dans le domaine juridique. Une forme nouvelle de fracture numérique entre les professionnels du droit pourrait s'opérer, les grands cabinets et directions juridiques bénéficiant désormais de systèmes d'IA hautement spécialisés, entraînés sur leurs propres bases de données, tandis que les petites structures devront se contenter d'outils généralistes, moins fiables, moins adaptées à la technicité du droit et plus exposées aux erreurs donc plus vulnérables aux « hallucinations » textuelles. Ce décalage dans la qualité des outils disponibles risque, à terme, d'accentuer les écarts de moyens entre praticiens et d'influer indirectement sur l'égalité des armes entre justiciables, selon la capacité de leurs conseils à mobiliser des systèmes d'analyse et d'argumentation assistés par l'IA.

---

[77] Décret n°2019-1089 du 25 octobre 2019 pris en application de l'article 4 de la loi n° 2019-222 du 23 mars 2019 de programmation 2018-2022 et de réforme pour la justice
[78] B. Chupin, R. Kaestlé, R. Sochon, « Médicys : la plate-forme de médiation des huissiers de justice », *Actu-juridique.fr*, 20 mai 2019, accessible sur : https://www.actu-juridique.fr/professions/medicys-la-plate-forme-de-mediation-des-huissiers-de-justice/, consulté le 4 mars 2023
[79] Voir en ligne le Centre d'arbitrage des affaires familiales dont l'offre est adossée à un outil qui se qualifie de « justice quantitative ».

54. **Vers une justice mondialisée en ligne ?** – Du point de vue d'auteurs comme le britannique Richard Susskind, le futur des tribunaux s'imagine aussi totalement dématérialisé et en ligne, organisé autour de standards mondiaux. Susskind propose ainsi l'avènement d'une plateforme *open source*, avec un ensemble de procédures intégrées adaptables aux différents types de juridiction[80]. Si la période de crise sanitaire a semblé donner une audience pour la vision de cet auteur[81], d'autres demeurent plus mesurées en cherchant à concilier les avantages du numérique et des impératifs procéduraux parfois difficilement transférables dans des environnements virtuels[82].

La question centrale qui se pose alors est celle des garanties des droits des usagers sur des plateformes dématérialisées, qui composent un tout nouvel écosystème gommant la délimitation entre justice étatique et justice privée. La polémique née aux Pays-Bas sur l'utilisation d'un système d'arbitrage en ligne « e-Court », du fait notamment du défaut d'information des usagers sur la portée de la procédure d'arbitrage et de la confusion entretenue avec une offre publique de justice, illustre les difficultés à faire encore cohabiter deux types d'organisation de traitement de contentieux sans méconnaître les dispositions posées par des cadres juridiques supranationaux tels que la Convention européenne des droits de l'homme, notamment au regard de ses articles 6, 8 et 13[83].

55. **Un accompagnement par l'éthique** – Afin de répondre aux enjeux particuliers posés par l'usage d'algorithmes et de « IA » dans un secteur aussi sensible que celui de la justice, la CEPEJ a adopté en décembre 2018 la toute première charte éthique européenne d'utilisation de l'IA dans les systèmes judiciaires et leur environnement[84]. Elle y définit cinq principes fondamentaux pour guider le développement et le déploiement des SIA dans la justice : 1) respect des droits fondamentaux à toutes les étapes de la vie d'un SIA, 2) non-discrimination, 3) transparence, 4) impartialité et intégrité, et 5) contrôle par l'utilisateur. Toutefois, comme tous les instruments de cette nature, la portée des principes est essentiellement déclaratoire et ne créent aucune obligation pour les concepteurs ou utilisateurs de systèmes d'IA.

56. **Une réglementation contraignante** – Depuis 2019, un nouvel environnement juridiquement contraignant s'est construit en Europe, avec l'adoption en 2024 d'un

---

règlement sur l'intelligence artificielle[85] (RIA) et d'une convention-cadre du Conseil de l'Europe[86]. S'ajoutant aux dispositions existantes comme le RGPD et la loi informatique et libertés, le RIA ambitionne de créer un cadre la confiance dans les différents usages des applications de « l'IA », en prévenant les atteintes sur les droits fondamentaux, la santé et la sécurité.

Certaines des applications relatives à l'administration de la justice vont ainsi se voir imposer un régime pour les applications à haut risque de « l'IA », avec des mécanismes de mise en conformité *ex ante* qui vont exiger des différents opérateurs une plus grande rigueur de conception et de documentation.

### III.2.  Des ambitions renouvelées pour un meilleur fonctionnement des tribunaux

### *III.2.1.  Le deuxième plan de transformation numérique*

57. **Un deuxième plan de transformation numérique pragmatique** – Quant au « backoffice » de la justice dans les tribunaux, le deuxième plan de transformation numérique présenté le 14 février 2023 illustre une réorientation pragmatique du ministère de la justice vers des réalisations réalistes et immédiatement utiles.

Ce plan s'inscrit en réponse du rapport d'étape de la Cour des comptes qui identifie certaines des causes du retard de la modernisation des outils numériques de la justice (insuffisance du renforcement de la fonction informatique du ministère, choix présentés comme contestables dans les priorités des projets et manque de suivi budgétaire[87]). La Commission des finances du Sénat est à l'origine de cette saisine de la Cour[88], après avoir dressé le constat, au cours de ses auditions sur le sujet de la crise sanitaire, que l'offre de service numérique n'avait pas permis aux juridictions de pouvoir investir rapidement le travail à distance. Ce constat concerne particulièrement l'informatique civile, dont la résilience aurait pu être meilleure[89].

58. **Un plan de transformation en six axes** – Ce nouveau plan a été élaboré en concertation avec les organisations syndicales, l'ensemble des directions du ministère, les représentants des partenaires institutionnels du ministère (avocats, commissaires de justice, notaires notamment), et les écoles d'application (ENM, ENG, ENPJJ, ENAP) est composé de six principaux axes :

---

[85] Règlement (UE) 2024/1689 du Parlement européen et du Conseil du 13 juin 2024 établissant des règles harmonisées concernant l'intelligence artificielle des règles harmonisées concernant l'intelligence artificielle (règlement sur l'intelligence artificielle)

[86] STCE n°225, Convention-cadre du Conseil de l'Europe sur l'intelligence artificielle et les droits de l'homme, la démocratie et l'État de droit, 17 mai 2024 (Convention cadre sur l'intelligence artificielle)

[87] Améliorer le fonctionnement de la justice – Point d'étape du plan de transformation numérique du ministère de la justice, Communication à la Commission des finances du Sénat, janvier 2022

[88] Sur les suites données au rapport de la Cour des comptes V. Rapport d'information fait au nom de la Commission des finances pour suite à donner à l'enquête de la Cour des comptes sur le plan de transformation numérique de la justice, accessible sur : https://www.senat.fr/rap/r21-402/r21-402_mono.html, consulté le 4 mars 2023

[89] Le retard du projet d'informatique civile Portalis, le cadre de sécurité à adapter pour un accès à distance aux logiciels actuels développés dans les années 1990 WinCi et ComCi, l'absence d'outils de mobilité sont quelques-uns des facteurs à évoquer.

- **Sécuriser et améliorer la qualité de l'existant pour redonner confiance aux agents :** l'objectif est de consolider et de fiabiliser les infrastructures (avec la poursuite du déploiement d'ordinateurs ultras portables, l'augmentation des débits sur le réseau informatique justice et l'installation de la fibre dans les 1 500 sites gérés par le ministère) ;
- **Remettre les utilisateurs et leurs pratiques au cœur de la transformation numérique :** les utilisateurs finaux seront au cœur des projets informatiques, en prenant mieux en compte leurs retours d'expérience et en les accompagnant de manière renforcée lors du déploiement de nouveaux logiciels ;
- **Accompagner les utilisateurs de façon globale :** le ministère souhaite accompagner les utilisateurs à travers une chaîne globale de soutien, sans leur imposer le transfert – aujourd'hui de fait – de la coordination de la résolution d'un problème complexe mettant en concours plusieurs services ; des informaticiens seront recrutés sur site pour cette fin, une identité numérique unique de connexion sera développée, ainsi que la signature électronique ;
- **Valoriser les données :** Outre l'évaluation des outils mis à la disposition des agents, cet axe concerne tant les capacités d'analyse (renfort de l'outil statistique) que décisionnelles (analyse et traitement des données) ;
- **Renforcer le réseau des partenaires de la justice grâce au numérique :** l'objectif est de travailler de manière plus étroite avec les professions, les écoles d'application et les associations ;
- **Garantir la sécurité, la résilience et la souveraineté :** dans un contexte d'augmentation de l'exposition aux risques de cyber-attaques en développant des services numériques et de pannes techniques considérées comme récurrentes par les utilisateurs, le ministère entend améliorer sa résilience en développant la redondance des réseaux dans les sites et en assurant lui-même la maintenance de son patrimoine.

Trois priorités ont été mises en avant dans la communication avec la presse : un objectif « zéro papier » d'ici 2027, l'arrivée de « vrais informaticiens » en juridiction et le lancement d'une application, portail avec les justiciables[90]. L'urgence a été de résorber la dette technique du ministère et l'obsolescence du patrimoine informatique du ministère, notamment par la refonte d'applicatifs imposant l'emploi de navigateurs obsolètes.

### *III.2.2. L'incubateur du ministère de la justice*

59. **Promouvoir l'innovation participative** – L'incubateur du ministère de la justice, déjà cité, soutient également de nombreuses initiatives émanant des agents publics, en mettant à disposition une méthodologie structurée et l'accompagnement d'experts (développeurs, designers, coachs produit, spécialistes métiers). Créé en 2021 avec le soutien de la DINUM, dans l'esprit des incubateurs publics tels que beta.gouv.fr et à la suite d'une première expérimentation d'un projet Justice au sein de cette structure, l'incubateur s'inscrit dans une vision claire : promouvoir une innovation participative « par et pour le terrain », en offrant aux agents, partenaires et justiciables un cadre

---

[90] A. Mestre, « Eric Dupond-Moretti présente son 'plan de transformation numérique' pour la justice », Le Monde, 14 février 2018

pour imaginer, tester et déployer rapidement des solutions adaptées aux irritants du quotidien.

Un premier appel à projets lancé début 2022 auprès de 90 000 agents a suscité près de 70 candidatures, dont 20 ont été retenues et présentées devant un jury. L'incubateur s'appuie sur un catalogue d'offre de services très structuré : immersion et recherche terrain, veille technologique, plateforme d'idéation, accompagnement méthodologique (approche produit, design, investigation), mise à disposition d'une force de réalisation (développeurs capables de créer des prototypes, POC technique ou d'usage, produit minimum viable en beta-production), ressources de prototypage, hébergement technique, financement allant de l'investigation jusqu'à la première version bêta, ainsi qu'une dynamique communautaire mêlant événements, retours d'expérience et capitalisation des connaissances.

60. **Les programmes spécifiques** – Au sein de cette dynamique, l'incubateur a développé plusieurs programmes spécifiques :

- les startups d'État, centrées sur la création de produits numériques à fort impact public selon les principes du manifeste beta.gouv (centrage utilisateur, impact mesurable, transparence, autonomie des équipes) ;
- les POC techniques et d'usage de l'ANJe, déjà cités, destinés à lever rapidement les risques techniques ou à évaluer l'appétence des utilisateurs ;
- les prototypes pour étendre à l'échelle nationale des initiatives locales prometteuses ;
- l'appui aux initiatives locales, afin de sécuriser la faisabilité technique et favoriser la généralisation de micro-solutions nées dans les juridictions ou services.

61. L'objectif est de soutenir le développement de « petites solutions à fort impact » tout en structurant et fédérant un véritable écosystème d'innovation interne, en lien avec d'autres incubateurs ministériels, les startups partenaires, et les acteurs de la communauté innovation Justice. Cette approche permet également de répondre aux attentes des directions métiers qui souhaitent davantage de structuration, de transparence, de coordination et de méthode dans le pilotage des projets innovants.

62. **Bilan en 2024** – En 2024, huit Startups d'État sont accompagnées conjointement par l'incubateur et beta.gouv.fr, portant par exemple sur le rappel automatique en ligne des convocations ou encore sur la gestion prévisionnelle des effectifs dans les juridictions[91]. Pour Mon Suivi Justice, le taux de non-présentation a été réduit de 30% avec l'envoi de relances des convocations par SMS. L'expérience a aussi révélé le manque de clarté des convocations envoyées, qui ont pu être résolues en phase d'investigation. L'incubateur bâtit également des prototypes d'outils, anime une communauté et produit du conseil métier.

---

[91] Il peut être cité pour l'ensemble du secteur justice, au 23 janvier 2024 : A-Just (https://beta.gouv.fr/startups/a-just.html) et Mon Suivi Justice (https://beta.gouv.fr/startups/justif.html) qui sont en phase d'accélération ; EXPERTS (https://beta.gouv.fr/startups/experts.html), infoParquet (https://beta.gouv.fr/startups/infoparquet.html), JAFER (https://beta.gouv.fr/startups/jafer.html), RDV MJD (https://beta.gouv.fr/startups/rdv.mjd.html), ROMIN (https://beta.gouv.fr/startups/romin.html) et Themis (https://beta.gouv.fr/startups/themis.html) qui sont en phase de construction.

L'incubateur joue ainsi un rôle de tremplin pour des innovations à fort potentiel, tout en contribuant à renouveler les pratiques professionnelles, diffuser une « culture produit » et renforcer la confiance dans la capacité du ministère à concevoir des solutions numériques utiles, rapides et centrées utilisateurs.

### III.2.3. La priorisation du déploiement de l'intelligence artificielle en 2025 au sein du ministère de la justice

63. **Une nouvelle priorité donnée au développement opérationnel de l'IA** – Alors que le second plan de transformation numérique ne priorisait pas le développement de l'intelligence artificielle[92], Gérald Darmanin, ministre de la justice, a souhaité donner une priorité à cette technologie en confiant au début 2025 une « mission d'accélération » de trois mois à un magistrat, Haffide Boulakras, ayant déjà contribué aux travaux des États généraux de la justice. Cette mission, regroupant dans un comité opérationnel toutes les composantes du ministère de la justice (services judiciaires, administration pénitentiaire, protection judiciaire de la jeunesse, administration centrale), a produit un rapport intitulé « L'IA au service de la justice : stratégie et solutions opérationnelles[93] » en mai 2025, remis officiellement en juillet 2025.

64. **Contenu du rapport** – Ce rapport propose une stratégie opérationnelle d'intégration de l'IA au sein du ministère, en réponse à un sentiment diffus de retard au regard de la dynamique d'adoption d'autres organisations, tant publiques que privées et aux attentes du terrain, comblées par l'utilisation peu sécurisée et non officielle de solutions commerciales.

Le document, riche de plusieurs annexes très opérationnelles, met en évidence l'ampleur des opportunités offertes par l'IA pour améliorer l'efficacité du service public, l'accessibilité au droit et les conditions de travail des agents. À partir d'une analyse systématique de soixante cas d'usage recueillis auprès des juridictions, de l'administration pénitentiaire, de la protection judiciaire de la jeunesse et des services d'administration centrale, la mission identifie des besoins convergents en matière de recherche, de synthèse, d'analyse de dossiers, de traitement des contentieux de masse et d'outillage des justiciables. L'étude souligne que les IA génératives constituent un levier décisif pour accélérer le traitement de l'information, harmoniser les pratiques, renforcer l'organisation interne et, plus largement, transformer les modes de travail de la chaîne judiciaire.

Le rapport insiste toutefois sur les prérequis juridiques, éthiques et techniques indispensables pour un déploiement responsable. L'encadrement actuel, résultant de la combinaison de textes (RGPD, directive Police-Justice, règlement sur l'IA, loi Informatique et Libertés), présente des zones d'incertitude, notamment concernant les finalités des traitements ou la qualification des systèmes à « haut risque ». Pour pallier ces difficultés, la mission propose des outils méthodologiques, en particulier un arbre décisionnel juridique, permettant aux concepteurs et utilisateurs de déterminer les règles applicables. Le rapport recommande également une évolution du cadre national, notamment la création d'espaces d'expérimentation (« bacs à sable réglementaires ») et la clarification des règles relatives à l'entraînement des modèles sur des données judiciaires sensibles. Sur le plan éthique, le rapport promeut l'instauration de principes directeurs et d'un label « IA digne de confiance en Justice », destiné à garantir transparence, sécurité, maîtrise humaine et prévention des biais.

---

[92] Cette technologie n'était pas pour autant ignorée, puisqu'en début 2025 le ministère avait développé une feuille de route avec quelques cas d'usages jugés comme prioritaire.

[93] Rapport L'IA au service de la justice : stratégie et solutions opérationnelles, *op.cit.*

65. **Le plan de travail proposé par le rapport** – La stratégie proposée repose sur un déploiement progressif articulé en trois phases. Dès 2025, il s'agit de mettre à disposition un assistant IA sécurisé et souverain pour l'ensemble des agents, capable d'assurer des tâches transversales (recherche, synthèse, rédaction, retranscription), et d'acquérir des solutions de recherche juridique augmentée par IA. Une douzaine de cas d'usage prioritaires doivent ensuite faire l'objet de développements dédiés entre 2026 et 2027, tandis que le ministère internalise les compétences techniques et installe des infrastructures souveraines d'hébergement. À partir de 2028, la consolidation de l'écosystème est envisagée : transfert complet vers les infrastructures internes, pilotage renforcé, publication de rapports annuels et poursuite de l'innovation.

Enfin, le rapport insiste sur l'importance de la gouvernance et de l'accompagnement humain. Il préconise la création d'une direction de programme dédiée à l'IA, adossée à un Observatoire de l'IA chargé du suivi éthique et stratégique. Il recommande également la mise en place d'un « campus du numérique » assurant une formation continue ambitieuse, afin de garantir l'appropriation des outils par les magistrats et agents. L'ensemble de ces mesures vise à faire de l'IA un instrument structurant de modernisation judiciaire, conciliant performance, souveraineté technologique et respect des principes fondamentaux de l'État de droit.

### III.2.4. L'appropriation de l'intelligence artificielle par la Cour de cassation

66. **Historique de l'appropriation de l'IA à la Cour de cassation** – La Cour de cassation fait partie des juridictions pionnières en France, et même probablement en Europe, en matière d'appropriation de l'intelligence artificielle.

Le développement de plusieurs initiatives employant de manière effective cette technologie s'est appuyé sur un très riche acquis en matière de données issu d'un long processus de dématérialisation amorcé dès les années 2000. Ce processus a permis à la juridiction suprême de se constituer un corpus numérique très substantiel (arrêts, mémoires, rapports, avis) ainsi que deux bases internes, JURINET et JURICA, pleinement opérationnelles au milieu des années 2010. À cette période, la Cour s'était également engagée dans le projet d'open data des décisions de justice[94] initié par la loi pour une République numérique de 2016, imposant la mise au point d'outils automatisés de pseudonymisation.

---

[94] *Cf. infra*

Les premières tentatives, fondées sur des systèmes de règles, ayant révélé leurs limites, la Cour s'est tournée vers l'apprentissage automatique au moment où ces techniques devenaient plus accessibles. C'est dans ce contexte qu'elle a participé, en 2018, au programme « Entrepreneurs d'intérêt général » (EIG), grâce auquel elle a constitué en 2019 une équipe interne de data scientists et de développeurs et a réalisé un prototype performant de pseudonymisation automatisée. Cette même année, le projet dénommé à l'époque « Open Justice » a marqué la première mise en œuvre réussie d'un moteur de pseudonymisation fondé sur de l'apprentissage automatique, puis l'outil d'annotation ergonomique « Label » a été conçu lors d'une seconde participation au programme EIG en 2021–2022.

67. **Une capacité interne d'innovation** – Ces développements ont ouvert la voie à une structuration progressive d'une véritable capacité interne d'innovation. Un algorithme d'orientation automatique des pourvois vers les chambres civiles a été mis en production peu après 2019, tandis qu'un partenariat de recherche avec l'Inria, retenu dans l'appel à projets « Manifestation d'intérêt IA » de 2019, a permis d'expérimenter des méthodes de détection des divergences de jurisprudence fondées sur la similarité textuelle et le résumé automatique.

Parallèlement, la mise en œuvre opérationnelle de l'open data s'est accélérée : dès 2023–2024, 48 tribunaux judiciaires publiaient leurs décisions civiles sur la plateforme Judilibre, avec un objectif d'exhaustivité pour l'ensemble des juridictions civiles à l'horizon fin 2025, tandis que la collecte des décisions prud'homales devait être achevée en septembre 2025, et que la diffusion des décisions pénales était annoncée pour 2026.

L'équipe interne, portée à une dizaine de membres en 2024–2025, a été intégrée au Service de documentation, des études et du rapport afin de maintenir une étroite articulation entre expertise technique et exigences juridiques.

68. **Une mission interne pour évaluer de nouveaux cas d'usage de l'IA** – Enfin, en 2024–2025, une mission interne confiée par le premier président et le procureur général a entrepris de recenser et d'évaluer de nouveaux cas d'usage de l'IA pour la Cour. Le rapport, remis et publié en avril 2025, analyse de manière approfondie les conditions dans lesquelles l'IA peut être intégrée, de manière éthique et maîtrisée, dans les activités de la Cour de cassation. Constitué d'un groupe de travail pluridisciplinaire rassemblant magistrats, greffe, juristes et data scientists, le projet s'appuie sur une méthodologie combinant recensement des besoins, auditions d'experts nationaux et internationaux, et étude de l'état de l'art en matière de technologies juridiques. La perspective adoptée est large : l'IA n'est pas réduite à sa seule dimension générative, mais appréhendée dans l'ensemble de ses déclinaisons (systèmes experts, apprentissage automatique, modèles hybrides), afin d'évaluer la pertinence réelle de ces technologies dans un contexte juridictionnel qui impose rigueur, fiabilité et transparence.

69. **Les critères d'analyse dégagés par le rapport de la mission interne** – Le rapport met en évidence une typologie structurée des critères d'évaluation applicables aux usages envisagés.

Trois grandes catégories de critères guident cette analyse : des critères éthiques, d'abord, plaçant au premier plan le respect des droits fondamentaux, la non-discrimination, la frugalité numérique et la préservation de la maîtrise humaine de la décision ; des critères juridiques, ensuite, centrés sur la conformité au Règlement IA de 2024 et au RGPD, notamment du fait du caractère sensible des données judiciaires ; enfin, des critères fonctionnels, techniques et économiques, qui visent à apprécier la faisabilité matérielle des projets, la disponibilité effective des données, l'état de l'art des modèles mobilisables et les coûts prévisibles d'entraînement, d'hébergement et d'exploitation des systèmes d'IA.

Le rapport souligne que cette grille d'analyse pourrait servir de référence pour d'autres juridictions.

70. **Les cinq familles de cas d'usage** – Sur cette base, cinq grandes familles de cas d'usage sont identifiées.

La Cour étudie d'abord des usages transversaux visant la structuration et l'enrichissement automatique des documents, puis des usages portant sur l'exploitation des écritures des parties, comme l'amélioration de l'orientation des pourvois ou la détection de connexités matérielles et intellectuelles. Une troisième famille concerne l'aide à la recherche et à l'exploitation des bases documentaires, incluant la recherche sémantique, des outils « RAG », la détection automatisée de divergences de jurisprudence ou encore l'analyse de litiges sériels. La dernière famille porte sur l'aide à la rédaction, depuis l'uniformisation des normes de style jusqu'à l'analyse de précédents pour certains contentieux.

Le rapport insiste toutefois sur une limite majeure : aucun usage d'aide à la décision n'a été jugé acceptable, en raison de l'exigence de préserver la plénitude de l'office du juge et d'éviter toute forme de dépendance ou de figement jurisprudentiel. Les limites intrinsèques des grands modèles de langage paraît également avoir motivé cette appréciation : il est illusoire de penser qu'une exposition à des millions de décisions en *open data* permettrait à un modèle de « comprendre » la justice car il n'accède qu'au produit final de la décision, non au déroulement contradictoire de l'audience, aux échanges écrits et oraux, ni à l'évolution du dossier qui entourent ce dispositif. Dans ces conditions, les modèles risquent de donner un poids excessif, pour des raisons purement statistiques, à des décisions de première instance nombreuses mais peu substantielles, au détriment d'arrêts de principe plus rares. Un tel « aplatissement » des sources par l'algorithme entre en tension avec la hiérarchie des normes et des autorités, qui constitue un élément central de la méthode juridique.

71. **Les principes directeurs de développement responsable de systèmes d'IA** – Enfin, le rapport dégage plusieurs principes directeurs pour un développement responsable des systèmes d'IA judiciaires. Le déploiement de ces outils suppose un socle informatique robuste, une maîtrise souveraine de l'hébergement et des données, ainsi qu'une gouvernance institutionnelle dédiée, comprenant un comité de suivi et un dispositif de formation continue pour les magistrats et personnels. Le rapport préconise également l'adoption d'une charte éthique interne, prolongeant la charte de la CEPEJ de 2018.

72. **L'approche pragmatique et raisonnée de la Cour de cassation** – Tout comme le rapport du ministère de la justice, l'ensemble des conclusions de ce rapport témoigne d'une approche dépassionnée, ni technophile ni technophobe : l'IA est présentée comme une opportunité majeure pour améliorer l'efficacité et la qualité du travail juridictionnel, à condition de rester strictement subordonnée à l'humain et de contribuer au renforcement, plutôt qu'à la dénaturation, de la mission normative de la Cour de cassation.

### III.3. La valorisation de la donnée : les ambitions de la justice française en matière d'*open data*[95]

### III.3.1. *Contexte de l'introduction des politiques d'open data des décisions de justice*

73. **Des injonctions parfois contradictoires** – La question de la valorisation des données judiciaires a pris, au cours des dernières années, une importance déterminante dans la réflexion autour de la modernisation de la justice française. L'ouverture des données publiques, dont les décisions de justice constituent sans doute la composante la plus sensible, s'inscrit dans un mouvement général d'ouverture, de transparence et de transformation numérique.

Toutefois, la mise en œuvre concrète de l'open data dans le champ juridictionnel ne se réduit pas à une simple opération technique de publication : elle mobilise des enjeux institutionnels, politiques, économiques, méthodologiques et éthiques d'une grande complexité.

L'ambition affichée depuis la loi pour une République numérique de 2016 est de mettre à disposition l'intégralité des décisions de justice sous forme de données réutilisables. Mais cette ambition rencontre aujourd'hui des obstacles considérables, qui révèlent les tensions profondes entre exigences de transparence et impératifs de protection, entre ouverture et souveraineté, entre innovation technologique et garanties fondamentales du procès équitable.

### III.3.2. *Historique de l'open data des décisions de justice : entre philosophie de la transparence et réalités judiciaires*

74. **Un projet politique et une politique publique** – L'expression open data renvoie simultanément à un projet politique et à une politique publique. Le terme recouvre à la fois un registre axiologique, reposant sur une idéologie d'ouverture et de transparence destinée à renforcer la démocratie, mais aussi un registre opérationnel, lié aux instruments concrets permettant de mettre les données à la disposition du public. La compréhension de cette dualité est essentielle, car l'open data judiciaire s'inscrit dans cette tension permanente entre idéaux politiques et contraintes techniques.

---

[95] Les développements relatifs à l'open data des décisions de justice résultent d'une exploitation de la littérature disponible ainsi que du séminaire sur la transformation numérique de la justice organisé les 12 et 13 novembre 2025 pour les étudiants du Master 2 Cyberjustice (disponible en ligne sur le site du Centre de culture numérique de Strasbourg).

75. **Les divergences historiques d'approche entre les modèles anglo-saxons et le modèle français** – Historiquement, l'ouverture des sources juridiques n'a pas suivi en France les modèles anglo-saxons, où la diffusion du droit repose depuis plus de trente ans sur des Legal Information Institutes tels que Cornell LII ou BAILII (British And Irish Legal Information Institute).

Aucun équivalent n'a émergé en France, où la diffusion du droit a longtemps été monopolisée par les éditeurs privés et régie par une logique de sélection opérée par les juridictions suprêmes. Ce rôle historique des cours suprêmes, dont les publications sélectives, à travers le Bulletin des arrêts de la Cour de cassation ou le Recueil Lebon du Conseil d'État, ont façonné une conception du droit fondée sur la hiérarchie des sources et la présomption de valeur doctrinale des décisions les plus hautes.

La sélection opérée par les juridictions suprêmes a structuré pendant près de deux siècles la culture juridique française. Les décisions des juridictions du fond, trop nombreuses, hétérogènes et souvent peu motivées, étaient jugées dépourvues d'intérêt pour la doctrine. Par ailleurs, une part importante de la collecte des décisions a longtemps été déléguée à un opérateur privé concessionnaire, qui ne diffusait qu'une fraction réduite des arrêts d'appel, le plus souvent sous forme de résumés. La décision brute était ainsi rarement accessible, difficile à obtenir et coûteuse.

La création de Légifrance en 1998 marque une première réponse étatique à ce déficit structurel. Mais, la couverture du service public de diffusion restait extrêmement faible à la veille de l'open data : seulement 0,3 % des décisions judiciaires et 9 % des décisions administratives étaient publiées. Cet écart entre les ambitions de transparence affichées et la réalité de la diffusion illustre l'ampleur du chantier qui s'ouvre au milieu des années 2010.

### III.3.3. Genèse des textes applicables à l'open data des décisions de justice

76. **La rupture de la loi pour une République numérique de 2016** – La véritable rupture normative intervient avec la loi pour une République numérique du 7 octobre 2016. L'inclusion des décisions de justice dans le champ de l'open data ne résulte ni d'une demande de la doctrine, ni d'une revendication des professions juridiques, ni d'un projet institutionnel porté par les magistrats ou les juridictions. Il s'agit d'un ajout tardif dans la navette parlementaire, largement motivé par un contexte économique particulier : l'émergence des legaltechs françaises, telles que Doctrine ou Predictice, et la volonté des pouvoirs publics d'encourager l'innovation dans le marché de l'information juridique. La loi impose néanmoins un principe général d'ouverture, inédit par son ampleur, plaçant l'État face à une exigence technique considérable : organiser la collecte, l'anonymisation, la structuration et la publication de plusieurs millions de décisions par an.

77. **La recherche d'équilibre de la loi de programmation de 2019** – La loi du 23 mars 2019 poursuit ce mouvement, tout en cherchant à protéger les professionnels de justice. Ce texte introduit des règles de pseudonymisation obligatoires pour les parties, maintient la visibilité des magistrats et greffiers, et crée un délit spécifique de profilage des professionnels à partir de l'analyse des décisions. Ces choix traduisent un équilibre complexe entre transparence et protection, et répondent notamment aux inquiétudes nées de l'affaire du billet Medium de 2016, dans lequel un auteur avait utilisé des statistiques rudimentaires pour critiquer individuellement des juges dans les contentieux d'OQTF[96].

Le cadre juridique actuel est codifié dans l'article L111-13 du Code de l'organisation judiciaire et du Code des relations entre le public et l'administration. Ce régime consacre une ouverture large et une liberté de réutilisation sous licence ouverte.

### III.3.4. Perspective critique de l'open data des décisions de justice : risques, controverses et reconfigurations contemporaines

78. **Les difficultés rencontrées par le cadre juridique de l'open data** – Si le cadre juridique de l'open data des décisions de justice paraît aujourd'hui établi, sa mise en œuvre se heurte à des difficultés. L'ouverture complète, initialement programmée pour 2025, accuse désormais plusieurs années de retard ; certaines catégories de décisions, notamment pénales, ne seront probablement pas publiées avant 2027 ou 2028.

79. **Des défis techniques** – Les obstacles rencontrés sont d'abord techniques : hétérogénéité des logiciels utilisés par les juridictions, absence de standardisation rédactionnelle, instabilité des systèmes d'information, difficultés persistantes de pseudonymisation automatique. Les juridictions du fond, où se produit la masse des décisions, constituent de fait le principal goulot d'étranglement du dispositif.

80. **Un écosystème reconfiguré par la banalisation de l'IA générative** – Les défis ne sont cependant pas seulement techniques. Les avancées de l'intelligence artificielle générative reconfigurent profondément l'écosystème. Les décisions de justice constituent un corpus incomplet, hétérogène et parfois pauvrement motivé, ce qui limite leur pertinence pour l'entraînement de modèles d'IA. Les modèles linguistiques donnent un poids statistique excessif aux décisions de première instance, en raison de leur nombre, et non de leur pertinence juridique, altérant potentiellement la compréhension de l'application du droit.

Par ailleurs, la réutilisation massive des données judiciaires par des acteurs privés, y compris non européens, nourrit une inquiétude croissante en matière de souveraineté. La gratuité des données françaises permet à des entreprises étrangères de constituer des corpus volumineux pour entraîner leurs modèles, sans que la France ne maîtrise l'usage qui en sera fait. Cette situation accentue la dépendance technologique et crée des distorsions de concurrence entre acteurs nationaux et internationaux.

---

[96] Supralegem, « L'impartialité de certains juges mise à mal par l'intelligence artificielle », *Medium*, 18 avril 2016, accessible sur : https://medium.com/@supralegem/l-impartialité-de-certains-juges-mise-à-mal-par-l-intelligence-artificielle-ee089170ddd3, consulté le 15 novembre 2025

La mise à disposition massive des décisions de justice comporte également un risque accru de profilage des magistrats, malgré l'incrimination pénale existant depuis 2019[97], et de « forum shopping », c'est-à-dire la recherche stratégique de juridictions perçues comme plus favorables.

81. **L'évolution de l'open data des décisions de justice : le rapport Ludet** – C'est dans ce contexte qu'intervient le rapport[98] du groupe de travail présidé par Daniel Ludet, conseiller honoraire de la Cour de cassation, qui propose une inflexion nette de la politique d'open data. Ce rapport envisage une réduction de l'ouverture initiale, à travers une pseudonymisation renforcée par défaut (tous les noms des magistrats et greffiers), une occultation plus large, et la mise en place d'un double circuit distinguant un accès public ne conservant que l'essentiel pour la compréhension juridique de la décision d'un accès à des flux plus intègres au gré à gré, réservé à certains usages et potentiellement payant.

L'ambition française en matière d'open data judiciaire s'inscrit donc dans une dynamique générale d'ouverture et de valorisation de la donnée publique. Cette ambition est portée par des objectifs multiples : démocratisation de l'accès à la justice, modernisation des pratiques professionnelles, stimulation de l'innovation technologique et développement d'outils d'intelligence artificielle juridiquement pertinents. Toutefois, cette ambition se trouve aujourd'hui confrontée à un ensemble de tensions qui en rendent la réalisation particulièrement complexe. Les obstacles techniques, institutionnels, juridiques et méthodologiques s'additionnent ; les inquiétudes liées à la protection des professionnels, à la vie privée et à la souveraineté s'intensifient ; les usages non encadrés de l'IA bouleversent un écosystème fragilisé.

Alors que la donnée judiciaire est devenue un enjeu stratégique de première importance, les mécanismes d'ouverture envisagés en 2016 apparaissent désormais insuffisants, parfois dépassés, parfois contestés. Le rapport Ludet illustre cette inflexion politique : l'ouverture ne constitue plus un horizon indiscuté, mais un objet de reconfiguration, voire de restriction. Dans ce paysage mouvant, la valorisation de la donnée judiciaire ne pourra pleinement se déployer qu'à la condition d'une gouvernance renouvelée et simplifiée, fondée sur la souveraineté, la sécurité, la transparence, mais également sur une appropriation méthodologique réelle par les juristes eux-mêmes. L'enjeu, désormais, n'est plus seulement d'ouvrir les données, mais de permettre leur usage maîtrisé, conforme aux principes fondamentaux du droit, et véritablement au service de la justice.

### III.3.5. Une initiative innovante en matière de valorisation de la données juridique et judiciaire : le Legal Data Space

82. **Genèse du Legal Data Space** – Née de la rencontre entre juristes, institutions publiques et acteurs européens de la legaltech, l'initiative Legal Data Space, lancée en mars 2025, s'inscrit dans un mouvement plus large de création d'espaces de données souverains en Europe. Portée notamment par Martin Bussy, consultant, et Thomas

---

[97] L. 111-13 du code de l'organisation judiciaire
[98] Rapport sur l'évolution de l'open data des décisions de justice, juillet 2025, accessible sur : https://www.justice.gouv.fr/sites/default/files/2025-08/rapport_oddj_2025.pdf, consulté le 15 novembre 2025

Saint-Aubin, juriste et entrepreneur ayant exercé au ministère de la Justice avant de rejoindre l'écosystème legal tech, cette initiative rassemble aujourd'hui plus de 140 partenaires publics et privés autour d'une ambition commune : créer le premier espace européen souverain dédié aux données juridiques. Le projet prend appui sur l'évolution du cadre réglementaire européen (RGPD, Data Governance Act, Data Act, AI Act) qui promeut une séparation des pouvoirs numériques et introduit la notion d'« intermédiaires de données », rôle précisément occupé par les data spaces.

83. **L'infrastructure du Legal Data Space** – Le Legal Data Space propose une infrastructure de partage sécurisé dans laquelle les membres conservent la maîtrise de leurs corpus tout en pouvant les rendre accessibles selon différents niveaux : usage strictement privé, partage au sein de « guildes professionnelles » ou mise à disposition sur une place de marché (« marketplace »).

La plateforme s'appuie sur des mécanismes d'authentification forte, de monétisation (« tokenisation ») et sur un corpus de règles (« rulebook ») qui encode nativement dans les systèmes d'information les règles issues du droit européen, de la *soft law* et des contrats de partage. Cette sorte de « codification algorithmique du droit » vise ainsi à rendre le droit calculable par la machine, dans une perspective de mise en conformité dès la conception (« compliance by design »). Les *data spaces* permettent ainsi de repenser le droit non seulement sous un angle textuel, mais également sous une forme algorithmique et statistique, ouvrant la voie à une reconfiguration de l'application du droit à l'ère numérique.

La plateforme permet aussi l'accès à une grande diversité de données juridiques qualifiées, favorise la création d'agents IA spécialisés, au-delà des simples agents conversationnels documentaires, et propose un cadre souverain permettant de prévenir les risques de « shadow IA » et de captation des données sensibles par des modèles propriétaires externes.

84. **La feuille de route du Legal Data Space** – La feuille de route du Legal Data Space combine déploiement technologique, structuration de la gouvernance et construction de communs numériques juridiques.

Elle vise notamment à renforcer l'interopérabilité via les standards européens, à développer des modèles d'IA fondés sur les référentiels, ontologies et méthodes du droit continental, et à associer durablement universités, ordres professionnels, think tanks et administrations.

À travers cette trajectoire, le Legal Data Space ambitionne de se positionner comme une infrastructure essentielle pour une souveraineté juridique européenne capable de soutenir une IA de confiance et un écosystème d'innovation aligné sur les principes fondamentaux du droit européen.

## III.4. Une transformation numérique de la justice n'ayant pas cédé au solutionnisme technologique mais dont les enjeux justifient encore une meilleure convergence entre les acteurs concernés

### III.4.1. Une transformation numérique de la justice ayant trouvé un équilibre d'approche avec l'intelligence artificielle

85. **Une transformation numérique n'ayant pas cédé au solutionnisme technologique** – À l'occasion de la 33e réunion plénière de la CEPEJ en 2019[99], le magistrat et universitaire français Jean-Paul Jean rappelait que les politiques publiques de la justice ne pouvaient être restreintes à l'amélioration de leur seule efficacité. Il concluait son propos en rappelant qu'il fallait rester humain et « pas trop technocrate » face aux transformations à venir, car la force de la justice, instrument de paix sociale, est avant tout sa qualité.

Parvenir à penser le numérique pour la justice en ce sens est certainement un défi considérable, en essayant de se départir d'un solutionnisme technologique très répandu dans les politiques publiques contemporaines, liant parfois de manière trop hâtive outils informatiques et progrès.

Particulièrement en matière de justice, le lien humain reste essentiel afin d'évaluer, contextualiser et individualiser les situations. La réduction de problèmes en objets manipulables par des algorithmes ne laisse pas nécessairement indemne la complexité des situations, la société ne pouvant pas aisément être mise en équation[100].

Les deux rapports sur l'IA produits par la Cour de cassation et le ministère de la justice en 2025 témoignent concrètement de cette prise de conscience et ouvrent des pistes d'appropriation de cette technologie avec une prise en compte approfondie des enjeux éthiques, juridiques et opérationnels.

### III.4.2. Les écueils persistants de la transformation numérique de la justice

86. **Une transformation numérique ne devant pas se tromper de cible** – Une autre question est la question de l'intermédiation entre les usagers et le service public de la justice. Le développement de services directement accessibles depuis des ordinateurs, des tablettes ou des téléphones intelligents ne se heurte pas qu'au problème de la fracture numérique. La qualité du conseil fourni en amont de la saisine d'une juridiction est essentielle pour coordonner la stratégie contentieuse. Un divorce, par exemple, ne génère pas qu'une procédure devant le juge aux affaires familiales : selon la complexité de l'affaire, d'autres problématiques peuvent émerger (comme des violences intrafamiliales, des contentieux immobiliers ou locatifs, etc).

Dès lors, le premier intermédiaire reste l'avocat, qui identifiera avec son client les différentes procédures, conviendra peut-être de recourir à une résolution amiable et l'accompagnera tout au long de son parcours. La transformation numérique de la justice ne s'adresse donc peut-être toujours aux bons utilisateurs finaux : moins que des portails ou des simulations en ligne s'adressant directement à une population

---

n'ayant pas toujours la culture juridique permettant de contextualiser son affaire, ce sont les avocats qui devraient peut-être bénéficier prioritairement d'outils adaptés pour fluidifier leurs relations avec les tribunaux.

87. **Une refonte de la gouvernance ministérielle bien engagée, mais restant à mener à son terme** – La transformation numérique des tribunaux a été, par ailleurs, freinée par plusieurs problématiques structurelles. La stratégie numérique n'a pas toujours été claire, stable et adossée à une vision rétrospective, prospective et comparée. Cette lacune se traduit par une difficulté à articuler la numérisation des procédures, l'évolution des métiers judiciaires et les finalités de la justice, ainsi qu'à concilier les besoins des professionnels et les attentes des usagers. Par ailleurs, la capacité du ministère à concevoir, piloter et maintenir son système d'information est demeurée trop longtemps limitée : outre une cartographie parfois incomplète de l'architecture des différents systèmes d'information, le ministère a géré sa dette technique sans avoir toujours les moyens de la résorber. La production informatique s'avère encore parfois fragile, notamment pour les plus gros systèmes, avec une incomplétude de la mesure globale et sectorielle de la qualité de service.

La faiblesse des ressources humaines affectées au numérique a été également longtemps chronique. Malgré les efforts réalisés durant le premier plan de transformation 2018-2022, les effectifs restent encore très inférieurs à la moyenne interministérielle, entraînant un niveau d'externalisation massif, y compris sur des fonctions stratégiques, et nuisant à la maîtrise interne des projets. À ces limites s'ajoutent un pilotage budgétaire qui a été perfectible[101] et une trajectoire financière 2023-2027 insuffisamment sécurisée. Enfin, la gouvernance du numérique, bien que renforcée récemment avec la création d'une direction du numérique (DNUM), demeure fragilisée par une relation entre maîtrise d'ouvrage (MOA) et maîtrise d'œuvre (MOE) historiquement clivée et par un déficit de maturité des directions métiers dans l'expression et la priorisation des besoins.

Au moment de la clôture de cette étude, la stratégie numérique continue de se clarifier en maintenant ses priorités, en intégrant des comparaisons internationales et en associant davantage les usagers[102]. Le renforcement des capacités de pilotage technique du ministère par une meilleure maîtrise de l'architecture et de la dette technique, ainsi que par la mise en place d'outils fiables de suivi de la qualité de service, est également en cours. Le renforcement et l'internalisation des compétences numériques constituent un axe essentiel, rendu concret par la création d'une direction de programme IA, de même que la sécurisation pluriannuelle des crédits alloués au deuxième plan de transformation numérique. La professionnalisation de la maîtrise d'ouvrage et l'approfondissement des mécanismes de priorisation et de gouvernance afin d'assurer la cohérence et la soutenabilité de la transformation engagée demeurent encore des défis.

88. **Le paradoxe de la promotion de modèles d'organisation libertariens pour porter des projets de l'État** – Les différentes générations d'administrations de l'informatique judiciaire se sont donc retrouvées trop souvent en tension entre des impératifs contradictoires et figées dans des macrostructures mélangeant contraintes réglementaires, bureaucratiques et politiques. La recherche d'agilité, sur le modèle de jeunes entreprises privées, afin de se départir des lourdeurs des circuits habituels de l'administration, s'est généralisé dans la fonction publique avec les « startups d'État » et il semble tout à fait positif que le ministère de la justice puisse également en tirer profit.

---

[101] V. rapport de la Cour des comptes « Améliorer le fonctionnement de la justice – point d'étape du plan de transformation numérique », janvier 2022

[102] Cf. la méthodologie adoptée par la mission « d'accélération » de l'IA conduite par le magistrat Haffide Boulakras.

Plusieurs types de difficultés se posent toutefois : outre le manque de maturité de certaines expérimentations et les succès restreint à des problématiques très sectorielles, il semble paradoxal que l'État trouve son salut dans la diffusion de méthodes d'entreprises dont l'idéologie libertarienne et ultra-individualiste vise à réduire toute forme d'ingérence étatique dans la gestion des affaires publiques. Ces dispositifs agrègent en réalité des compétences dans des cadres contractuels parfois très précaires, ne garantissant pas à l'État de pouvoir disposer de manière stable de compétences de haut niveau.

89. **Les rapports asymétriques avec le secteur privé** – Les partenariats avec le secteur privé pour dynamiser l'offre publique présentent également un certain nombre de difficultés, dont certaines ont été déjà plutôt bien documentées, notamment par la Cour des comptes. La sous-traitance massive des dernières décennies a révélé combien le pilotage de prestations exigeait un haut niveau d'expertise du côté du secteur public[103]. À défaut, les dérives financières, notamment en ce qui concerne l'assistance à maîtrise d'ouvrage (AMOA), se sont révélées d'un niveau considérable[104].

Des partenariats technologiques, où l'État fournirait des données et une entreprise les traiteraient pour en produire une valeur ajoutée, se révèlent également tout à fait asymétriques[105]. Le modèle de l'*open data* est à ce titre éloquent : présenté comme un accélérateur de développement économique, la contrepartie pour l'intérêt général peine à être mesurée[106]. Les licences à envisager pour exploiter massivement des données publiques doivent garder à l'esprit les modèles extractifs d'entreprises comme Google ou Meta/Facebook : alors que la philosophie de l'*open data* a pour objectif d'accroître les communs, la réexploitation de données publiques et l'élaboration de modèles avec de l'apprentissage automatique sans contrepartie pour l'intérêt général paraît devoir être soigneusement reconsidérée quant aux réelles finalités poursuivies.

90. **Une politisation du sujet ne permettant de s'inscrire dans un temps long** – La politisation de la transformation numérique de la justice, avec des ministres de la justice imprimant chacun leurs urgences et priorités, a été inévitable au vu du soutien de haut niveau nécessaire pour soutenir l'impact sur les finances publiques et des enjeux de gouvernance d'une technologie si structurante. Mais cette même politisation a probablement été, dans le même temps, un frein pour envisager le lancement des réorganisations d'ampleur, au-delà du temps des mandats politiques, comme celle des investissements dans les infrastructures, ou encore a alimenté les concurrences entre services.

---

[103] Rapport conjoint Inspection des finances / Conseil général de l'économie, de l'industrie, de l'énergie et des technologies, le pilotage et l'audit des grands projets informatiques de l'État, mars 2012
[104] Communication de la Cour des comptes à la Commission des finances du Sénat, La conduite des grands projets numériques de l'État, juillet 2020
[105] V. notamment l'expérience des premiers essais de pseudonymisation des décisions par la Cour de cassation en lien avec la société Doctrine
[106] V. par exemple S. Goëta, C. Mabi, « L'open data peut-il (encore) servir les citoyens ? », *Mouvements*, vol. 79, no. 3, 2014, pp. 81-91 ou encore E. Barthe, « Open data : le désenchantement », *Blog precisement.org*, juillet 2021, accessible sur : https://www.precisement.org/blog/Open-data-le-desenchantement.html

### III.4.3. Construire un nouveau récit de la transformation numérique de la justice

91. **Des attentes fortes justifiant des évolutions concrètes et perceptibles** – En conclusion, malgré les progrès indéniables, les attentes demeurent encore fortes dans les juridictions et auprès des professionnels. Le changement de perception ne pourra naître que d'évolutions concrètes et perceptibles, nourrissant l'écriture d'un nouveau récit partagé.

Ainsi, le chantier de la dématérialisation serait à accompagner d'outils simples d'authentification des documents (signature électroniques, blockchains) dont il semble de plus en plus difficile de justifier l'absence en 2025. Le stockage sécurisé et disponible de ces documents sera un autre défi, tout comme celui des infrastructures pour supporter de fortes volumétries d'échanges.

Celui du développement de l'emploi de « l'IA » devra affronter les risques de déception des solutions qui seront mises à disposition des personnels : sous-performance par rapport aux offres du commerce, ajout d'une interface supplémentaire dans des processus de travail déjà trop divertis par de multiples outils non connectés entre eux, incertitude du cadre réglementaire, voici quelques-uns des écueils que la direction de programme IA du ministère devra affronter. Il devrait peut-être également être distingué ce qui devra continuer à être externalisé, comme l'entraînement initial des modèles, et ce qui doit impérativement rester interne à l'institution judiciaire, comme l'utilisation sur des dossiers individuels ou des procédures protégées par le secret. La piste d'une exécution locale de modèles adaptés à l'activité judiciaire (c'est-à-dire directement sur le micro-ordinateur de l'utilisateur) plutôt que dans le cloud, même si cela implique des performances techniques moindres, offrirait des garanties plus adéquates en matière de confidentialité, de traçabilité et de maîtrise institutionnelle des flux.

La justice manipulant des données parmi les plus sensibles sur les citoyens, il en découle que les infrastructures utilisées doivent être pleinement maîtrisées par l'État, excluant le recours à des solutions numériques non souveraines. La question de la souveraineté numérique devient ainsi un enjeu central, non seulement technique, mais institutionnel et démocratique.

92. **Des attentes fortes justifiant d'imaginer de nouvelles structures à disposition des parties prenantes** – Enfin, aller au-delà de la simple approche instrumentale, en réaction à diverses pressions (attentes des utilisateurs, état du marché, stratégies politiques) et concevoir une vision prospective de ce que devrait être une justice dans notre nouvel environnement socio-technique sont des impératifs qui devraient traités en amont de toute définition de stratégie en la matière. Cesser de confondre finalités et moyens ne pourra résulter que d'une réflexion approfondie, multi-disciplinaire et dégagée du temps court politique.

Il faudrait donc probablement compléter l'outillage à disposition de l'ensemble des acteurs de la justice et du droit de structures favorisant l'expérimentation dans le temps court et la réflexion dans le temps long.

En France (voire en Europe), il manque probablement des tiers-lieux originaux, à l'image du laboratoire de cyberjustice de Montréal pour penser et agir à la fois, en mettant en relation les différentes parties prenantes dans une organisation souple, réactive et soucieuse de l'intérêt général, sans dépendance forte à des impératifs poltiques ou économiques du secteur privé, qui s'enchevêtrent même parfois et imposent des choix technologiques peu opportuns.

**Annexe 1 – Professionnels étant intervenus auprès des étudiants du M2 Cyberjustice (2022-2025)**

Le directeur scientifique et les étudiants remercient chaleureusement les 24 intervenants suivants de leur aimable contribution à ces travaux universitaires de documentation de la transformation numérique de la justice.

- Magued Abdel Maaboud, sous-directeur applications, innovations et gouvernance, secrétariat général du ministère de la justice – L'innovation au ministère de la justice, 23 janvier 2024

- Jean-François Beynel, premier président de la Cour d'appel de Versailles – Vues sur la transformation numérique de la justice, 10 novembre 2022

- Camille Bordere, maîtresse de conférences – Perspective critique de l'open data, 13 novembre 2025

- Haffide Boulakras, directeur-adjoint de l'École Nationale de la magistrature – Retour d'expérience PPN et transformation numérique de la justice, 18 janvier 2024

- Thomas Cassuto, président de chambre à la Cour d'appel de Reims – Vues sur la transformation numérique de la justice, 23 janvier 2025

- Olivier Chevet, magistrat, responsable d'études et de recherche à l'Institut Robert Badinter – La donnée : « carburant » de la transformation numérique et de l'intelligence artificielle, 13 novembre 2025

- Romain Cousin, directeur stratégie produits, Groupe Karnov – Historique et actualité du traitement de la donnée juridique et judiciaire, 13 novembre 2025

- Steeve Delor, Chief Digital Officer, sous-directeur adjoint organisation judiciaire et innovation à la direction des services judiciaires, ministère de la justice – L'accompagnement des juridictions judiciaires dans leur transformation numérique, 23 janvier 2024

- Delphine Deneubourg, cheffe du bureau IFN (Innovation et Fabrique numérique) – L'incubateur du ministère de la justice, 21 novembre 2024

- Jean-Bernard Desjardins, directeur du secrétariat du parquet du tribunal judiciaire de Paris – Présentation de l'informatisation de la chaîne pénale (micro et mini-pénale, NCP), 29 septembre 2022

- Christian Elek, magistrat honoraire, ancien directeur du Casier judiciaire national – Le Casier judiciaire national, 29 septembre 2022 et 23 janvier 2024

- Harold Épineuse, directeur-adjoint de l'Institut Robert Badinter – Introduction à la transformation numérique de la justice et conférence conclusive, 30 septembre 2022, 18 février 2023,  25 janvier 2024 et 12 novembre 2025

- Thierry Ghera, président du TJ de Strasbourg – Informatisation de la chaîne civile (RPVA, OPALEX, IPWeb), 29 septembre 2022 et 23 janvier 2024

- Aurore Hyde, professeure agrégée – Réglementation et régulation de l'IA en justice, 12 novembre 2025

- Tarik Lakssimi, professeur agrégé et directeur de la recherche de l'ENM – Les enjeux éthiques de la transformation numérique de la justice, 12 novembre 2025

- Sylvie Mergès, inspectrice de la justice – L'informatisation de la chaîne civile (1995-2000) et réflexions sur la transformation numérique de la justice, 29 septembre 2022

- Jean-Pierre Poussin, magistrat honoraire, ancien délégué de la COMIRCE – Histoire et des activités de la COMIRCE, 23 septembre 2022

- Matthieu Quiniou, avocat et expert international – Approche internationale et européenne de la cyberjustice, 12 novembre 2025

- Edouard Rottier, chef de projet Open Data, Cour de cassation – Emploi de l'intelligence artificielle à la Cour de cassation, 6 décembre 2024

- Thomas Saint-Aubin, directeur de l'innovation @MyLegiTech, coordinateur Legal data space – Avenir de la donnée juridique et judiciaire : l'exemple du Legal data space, 13 novembre 2025

- Thibaut Spriet, juge des contentieux de la protection au TJ de Rennes – Présentation d'une application de gestion du contentieux de la consommation, 21 octobre 2022 et 23 janvier 2024

- Claire Strugala, chargée de mission auprès de la cheffe du service de l'expertise et de la modernisation – Vues sur l'appropriation et le développement de l'IA par le ministère de la justice, 21 novembre 2024

- Haï-Ha Trinh-Vu, chef du pôle Lab Data Justice – La capitalisation de la donnée au service de l'IA des administrations de la justice, 13 novembre 2025

- Jessica Vonderscher, procureure de la République près le TJ de Belfort – Présentation de l'agence nationale du TIG et du projet de dématérialisation au sein du parquet de Belfort, 19 octobre 2022

Les interventions des professionnels cités ont été réalisées en leur capacité personnelle, au regard de leur expérience passée ou présente d'informatisation et de numérisation des services judiciaires. Elles ne traduisent ou ne reflètent aucune position officielle de leurs institutions d'appartenance.

Les transcriptions et les résumés de leurs interventions, à des fins de recherches ultérieures, ont été remis à l'Institut Robert Badinter (IRB).

**Annexe 2 – Étudiants du M2 Cyberjustice ayant contribué à la documentation de la présente étude**

*Promotion 2022-2023*

Sous-groupe ayant traité des aspects relatifs à l'histoire de l'informatique judiciaire :

- Cassiopée Bal
- Lucia Berdeil
- Chloe Borde
- Lise Bujon
- Christiane Chiotis
- Natalie Chiotis
- Emma Fritz
- Manon Perbal
- Hannah Siegrist
- Léo Tarpin
- Lilian Vasseur
- Fiona Vercelli

Sous-groupe ayant traité des aspects relatifs à la transformation numérique contemporaine de la justice :

- Eda Alkin
- Jeanne Gotié
- Julie Gracia
- Maud Igersheim
- Hanan Kaichouh
- Chloé Padar
- Delija Talevic
- Jeanne Thielges
- Ugo Volto-Garoscio

*Promotion 2023-2024*

Sous-groupe ayant traité des aspects relatifs à l'histoire de l'informatique judiciaire :

- Lucie Speyser
- Amandine Derouet
- Ameni El Amri
- Olivia Martin
- Emma Wack Wendling
- Thomas Busser
- Antoine Candau
- Emilya Ramazanova
- Gohar Simonyan
- Roxana Vener

Sous-groupe ayant traité des aspects relatifs à la transformation numérique contemporaine de la justice :

- Justine Van Bever

- Dunice Daigrier
- Jeanne Casamatta
- Noor Rafouk
- Julie Freiermuth
- Léane Kastner
- Kenza Ghezloun
- Léonard Simoens
- Matthias Levieux
- Ai Pei Choo

### Promotion 2024-2025

Sous-groupe ayant traité des aspects relatifs à l'histoire de l'informatique judiciaire :

- Joao Pedro Alcantara
- Raphael Ormezzano
- Juliette Pons
- Justine Denner
- Siryne Mrbati
- Murielle Moussa
- Nina Ghanbary
- Clara Bonnard
- Clara Castillon
- Laeticia Eschlimann

Sous-groupe ayant traité des aspects relatifs à la transformation numérique contemporaine de la justice :

- Julia Berthaud
- Louis Perrin-Conrad
- Julie Branco de Vera
- Camille Flamand
- Nadia Benyahia
- Liza Samaha
- Gohar Simonyan
- Lea Rozzo
- Justine Foos

### Promotion 2025-2026

Finalisation et relecture de l'étude :

- Camille Alexandre
- Océan Amr-Tardif
- Lorean Barrera Santana
- Basma Boudraa
- Monica Burgos Herrera
- Yousra El Barbary
- Jeanne Fages

- Ayse Gulcan
- Arthus Jallade
- Camille Joly
- Camille Maulino
- Clara Michaud
- Inès Noë
- Louise Passeri
- Salomé Petrozzi
- Louise Rasamizatovo
- Ilona Ribeiro
- Melina Smailovic

## Annexe 3 – Bibliographie sélective

### *Instruments européens*

STCE n°225, Convention-cadre du Conseil de l'Europe sur l'intelligence artificielle et les droits de l'homme, la démocratie et l'État de droit, 17 mai 2024 (Convention cadre sur l'intelligence artificielle)

Règlement (UE) 2024/1689 du Parlement européen et du Conseil du 13 juin 2024 établissant des règles harmonisées concernant l'intelligence artificielle des règles harmonisées concernant l'intelligence artificielle et modifiant les règlements (CE) n°300/2008, (UE) n°167/2013, (UE) n°168/2013, (UE) 2018/858, (UE) 2018/1139 et (UE) 2019/2144 et les directives 2014/90/UE, (UE) 2016/797 et (UE) 2020/1828 (règlement sur l'intelligence artificielle ou AI Act)

Conseil de l'Europe, Systèmes judiciaires européens, efficacité et qualité de la justice – Rapport thématique : l'utilisation des technologies de l'information dans les tribunaux en Europe, Études de la CEPEJ n°24, décembre 2016

Conseil de l'Europe, Lignes directrices sur la conduite du changement vers la Cyberjustice, CEPEJ, 2017

Conseil de l'Europe, CEPEJ(2018)14, Charte éthique européenne sur l'utilisation de l'intelligence artificielle dans les systèmes judiciaires et leur environnement, CEPEJ, décembre 2018

Conseil de l'Europe, Boîte à outils pour soutenir la mise en œuvre des Lignes Directrices sur la conduite du changement vers la cyberjustice, CEPEJ, 2019

UNESCO, UNESCO Global Judges' Initiative: survey on the use of AI systems by judicial operators, 2024

UNESCO, Manuel de formation mondial : l'IA et l'état de droit pour le pouvoir judiciaire, 2024

### *Rapports*

Rapport de la mission d'information d'Etienne Blanc relatif aux carences de l'application des peines et à l'évaluation de l'application Cassiopée – Rapport n°3177 déposé à la présidence de l'Assemblée nationale le 16 février 2011

Rapport conjoint Inspection des finances / Conseil général de l'économie, de l'industrie, de l'énergie et des technologies, le pilotage et l'audit des grands projets informatiques de l'État, mars 2012

Institut Montaigne « Justice : Faites entrer le numérique », novembre 2017

Rapport des chantiers de la justice « Transformation numérique », janvier 2018

Communication de la Cour des comptes à la Commission des finances du Sénat, La conduite des grands projets numériques de l'État, juillet 2020

Annexe 25, Rapport sur le numérique du comité de pilotage des États généraux de la justice, 17 mars 2022

Communication de la Cour des comptes à la Commission des finances du Sénat, Améliorer le fonctionnement de la justice – Point d'étape du plan de transformation numérique du ministère de la Justice, janvier 2022

Inspection générale de la justice, Inspection générale des finances, Conseil général de l'économie, de l'industrie, de l'énergie et des technologies, La transformation numérique de la justice, mai 2023 (non publié)

Sénat, Rapport d'information n°216 enregistré à la Présidence du Sénat le 18 décembre 2024, fait au nom de la commission des lois constitutionnelles, de législation, du suffrage universel, du Règlement et d'administration générale sur l'intelligence artificielle et les professions du droit, par M. Christophe-André FRASSA et Mme Marie-Pierre de LA GONTRIE

Cour de cassation, Préparer la Cour de cassation de demain : Cour de cassation et intelligence artificielle, avril 2025

Ministère de la justice, L'IA au service de la justice : stratégie et solutions opérationnelles, mai 2025

Ministère de la justice, L'évolution de l'open data des décisions de justice, juillet 2025

### *Ouvrages, articles et thèses*